\documentclass[journal,onecolumn,draftclsnofoot,11pt]{IEEEtran}
\usepackage{graphicx,cite,amsmath,amssymb,autobreak,hhline,booktabs,stfloats,bm,amsthm,hyperref,times,subfigure,multirow,algpseudocode,}
\usepackage{algorithm,algpseudocode,multicol,threeparttable,dcolumn,latexsym,rotating,textcomp,colortbl,enumerate}
\usepackage{verbatim}
\usepackage{caption}
\usepackage[top=2cm, bottom=2cm, left=2cm, right=2cm]{geometry}
\usepackage{graphicx}
\usepackage{grffile}
\usepackage{stfloats}
\begin{document}

\newtheorem{lemma}{\bf Lemma }
\newtheorem{theorem}{\bf Theorem}
\newtheorem{corollary}{Corollary}
\renewcommand{\algorithmicrequire}{\textbf{Input:}} 
\renewcommand{\algorithmicensure}{\textbf{Output:}}
\newtheorem{remark}{Remark}
\title{Toward Secure and Private Over-the-Air Federated Learning}

\author{Na Yan, Kezhi Wang, Kangda Zhi, Cunhua Pan, Kok Keong Chai \\ and H. Vincent Poor, \emph{Life Fellow, IEEE} \thanks{This work of Na Yan was supported by China Scholarship Council. \itshape (Corresponding author: Kezhi Wang and Cunhua Pan.)\upshape
		
		Na Yan, Kangda Zhi and Kok Keong Chai are with the School of Electronic Engineering and Computer Science, Queen Mary University of London, London, E1 4NS, U.K. (e-mail: n.yan, k.zhi, michael.chai@qmul.ac.uk).	
		
		Kezhi Wang is with the Department of Computer and Information Sciences, Northumbria University, NE1 8ST,  U.K. (e-mail: kezhi.wang@northumbria.ac.uk).
		
		Cunhua Pan is with the National Mobile Communications Research Laboratory, Southeast University, Nanjing 210096, China. (email: cpan@seu.edu.cn).
	
	
	


H. Vincent Poor is with the Department of Electrical and Computer Engineering, Princeton University, Princeton, NJ 08544, USA. (email: poor@princeton.edu)}
}
\maketitle
\begin{abstract}
In this paper, a novel secure and private over-the-air federated learning (SP-OTA-FL) framework is studied where noise is employed to protect data privacy and system security. Specifically, the privacy leakage of user data and the security level of the system are measured by differential privacy (DP) and mean square error security (MSE-security), respectively. To mitigate the impact of noise on learning accuracy, we propose a channel-weighted post-processing (CWPP) mechanism, which assigns a smaller weight to the gradient of the device with poor channel conditions. Furthermore, employing CWPP can avoid the issue that the signal-to-noise ratio (SNR) of the overall system is limited by the device with the worst channel condition in aligned over-the-air federated learning (OTA-FL). We theoretically analyze the effect of noise on privacy and security protection and also illustrate the adverse impact of noise on learning performance by conducting convergence analysis.
Based on these analytical results, we propose device scheduling policies considering privacy and security protection in different cases of channel noise. In particular, we formulate an integer nonlinear fractional programming problem aiming to minimize the negative impact of noise on the learning process. We obtain the closed-form solution to the optimization problem when the model is with high dimension. For the general case, we propose a secure and private algorithm (SPA) based on the branch-and-bound (BnB) method, which can obtain an optimal solution with low complexity. The effectiveness of the proposed CWPP mechanism and the policies for device selection are validated through simulations.

\centerline{\textbf{Intex Terms} } 

Federated learning, differential privacy, mean square error security, branch-and-bound, integer nonlinear fractional programming.
\end{abstract}

\section{Introduction}
The vast amount of valuable data generated at the edge of wireless networks has enabled various artificial intelligence (AI) services for end-users by exploiting deep learning \cite{lecun2015deep}. In most applications, such as the internet of things (IoT), unmanned aerial vehicles (UAVs), or extended reality (XR), data from sensors normally needs to be constantly collected and processed. There are numerous machine learning (ML) algorithms that have been developed to leverage these large-scale datasets. Most of the conventional ML algorithms are centralized, which aggregates all the raw data to a powerful central server where models are trained \cite{YANG202233}. However, such centralized ML approaches may become increasingly undesirable as privacy concerns and the size of the dataset increase. Specifically, uploading large amounts of raw data from the edge of the network to the centre is generally not feasible due to latency, bandwidth or power constraints and, most importantly, it may directly expose personal information.

To overcome these challenges, federated learning (FL) \cite{mcmahan2017communication} has been proposed as a privacy-enhancing distributed ML technique, which enables devices to train a model in a decentralized manner with the help of a central controller, such as a base station (BS). Specifically, edge devices firstly download the latest global model parameter from the BS and then compute gradients or update model parameters locally based on local datasets. Then, the gradients or updated model parameters are sent to the BS for the global model update. By training models locally, FL not only makes full use of the computing power of the edge devices, but also effectively reduces the power consumption, latency, and privacy exposure due to the transmission of raw data. In practice, FL as a promising technique has been widely applied to several systems, such as model training in wireless networks \cite{zhu2020toward} and content recommendations for smartphones \cite{bonawitz2019towards}. 

Inspired by the high communication efficiency of over-the-air computation (AirComp) \cite{nazer2007computation,zhu2019broadband,yang2020federated,amiri2020machine}, over-the-air FL (OTA-FL) has been proposed and attracted a great deal of attention. OTA-FL schedules devices concurrently uploading their local models or gradients through a wireless multiple-access channel (MAC). The gradients in prevailing OTA-FL are normally transmitted via an analog transmission. In this way, the BS receives an aggregated gradient or model directly thanks to the waveform-superposition property. The number of dimensions used for transmitting the gradients or models is independent of the number of devices, which makes it highly energy and bandwidth efficient compared with the traditional communication-and-computation separation method, especially when the number of devices is large \cite{nazer2007computation,goldenbaum2013harnessing}.




Although FL offers basic privacy protection, which benefits from the fact that all raw data is processed locally, it is far from sufficient. On the one hand, some studies revealed that various attacks \cite{fredrikson2015model,melis2019exploiting,shokri2017membership} can infer the individual data or recover part of training data by attacking the exchanged messages, i.e., the gradients and models \cite{nasr2018comprehensive}. This is because the updated model or gradient is obtained based on local data and therefore may contain some information from raw data \cite{song2017machine}.
Specifically, if the BS is ``honest but curious", it may silently infer the private information from the intermediate gradients or the trained models.
One of the countermeasures to prevent such privacy leakage of FL is differential privacy (DP) \cite{dwork2014algorithmic} which randomizes the disclosed statistics. Some studies focused on differentially private OTA-FL. In \cite{seif2020wireless}, artificial Gaussian noise was added to each gradient before transmitting if channel noise cannot provide sufficient privacy protection, and a static power allocation scheme was proposed to determine the scale of the artificial noise. Instead of introducing artificial noise, the work of \cite{koda2020differentially} proposed a more energy-efficient strategy to guarantee DP by adjusting the transmit power. The authors of \cite{liu2020privacy} investigated differentially private FL in both orthogonal multiple access (OMA) and non-orthogonal multiple access (NOMA) channels and proposed adaptive power allocation schemes. Nevertheless, similar to the most prevailing OTA-FL studies, all the above works considered an aligned aggregation by controlling transmit power. Although gradient alignment can ensure an unbiased gradient estimation at the BS, the signal-to-noise ratio (SNR) of the system will be limited to a pretty low level if any device suffers from a poor channel condition.



On the other hand, the broadcast nature of wireless channels makes FL vulnerable to security attacks. Although security and privacy are used interchangeably in the existing literature, it is important to highlight the distinction between them. In particular, privacy concerns normally refer to the disclosure of personal information from open access data, while security concerns refer to unauthorised access or alteration of data \cite{ma2020safeguarding}. Eavesdropping on the FL communication channels is often the first step for malicious third parties to launch security attacks. The main difference between eavesdropping attack and privacy issue at the BS is that the BS is authorised to receive the gradients and updated models while the eavesdropper is prohibited. To enhance the security of FL communications, the authors of \cite{van2021latency,xie2022securing} adopted a covert communication (CC) technique with which a friendly jammer transmits jamming signals to prevent an eavesdropper from detecting the update transmission of the local model from mobile devices in FL. The work of \cite{yao2021secure} utilized power control to improve the security of FL in the internet of drones (IoD) networks where security rate was employed to measure the security of wireless communications. However, all of these existing works on FL security investigated how digitally coded communication rounds can be protected against eavesdropping. To the best of our knowledge, the security of analog OTA-FL over a wiretap channel has not yet been considered.


\subsection{Contributions}
Inspired by these research gaps, in this paper, we propose a secure and private OTA-FL (SP-OTA-FL) framework where we try to utilize noise to protect privacy and security which are measured by DP and mean square error security (MSE-security) \cite{frey2020towards}, respectively. To mitigate the impact of noise on learning accuracy, we propose a channel-weighted post-processing (CWPP) mechanism, which does not require gradient alignment. We also conduct the privacy, security and convergence analysis to illustrate the impact of noise on privacy, security protection and learning performance. In addition, we propose different device selection policies for different cases of channel noise. Particularly, we formulate an integer nonlinear fractional optimization problem to minimize the optimality gap while guaranteeing privacy and security in the case of insufficient channel noise. 
The closed-form solution to the optimization problem is obtained when the model is with high dimension.
Based on the branch-and-bound (BnB) algorithm, a secure and private scheduling algorithm (SPA) with low complexity is proposed for the general case. The proposed post-processing mechanism and the device selection policies are validated through simulation. Our main contributions are summarized as follows.

\begin{itemize}
	\item We firstly propose a novel SP-OTA-FL framework where noise is employed to protect privacy and security. DP and MSE-security are used to measure the privacy leakage and the security level of the OTA-FL system, respectively. In the proposed framework, a part of the devices are selected as uploaders to participate in the training process and some other devices are selected as jammers to send Gaussian artificial noise aiming at deteriorating the eavesdropper’s SNR so as to guarantee the security of the system and privacy of user data at the BS.  
	To the best of our knowledge, this is the first work to consider privacy and security issues together in an FL system. More importantly, this is the first to focus on the security of analog OTA-FL, where gradients are transmitted in an uncoded way and therefore, are more vulnerable to security threats.
	\item We propose a CWPP scheme to alleviate the negative impact of noise on the utility of aggregated gradient, which could improve learning performance. By employing the CWPP mechanism, the gradients are no longer forced to be aligned during the transmission, and therefore, can avoid the issue that the SNR is limited by the device with the worst channel condition.
	\item The privacy, security and convergence analysis are conducted. The results theoretically reveal the benefits of noise on privacy and security protection as well as the negative impact on learning performance. The convergence analysis shows that the proposed SP-OTA-FL can converge with the rate of $\mathcal{O} \left( \frac{1}{T} \right)$. In particular, if the data is independently and identically distributed (IID), the optimality gap tends to zero as the number of data samples and training rounds grows.
	\item  We analyze the decision of device selection in three cases where: (1) channel noise is sufficient for protecting privacy and security with all device participation; (2) channel noise is sufficient for protecting privacy and security with partial device participation; (3) channel noise is insufficient for protecting privacy and security with any device participation. We propose two policies to schedule devices to ensure user privacy and system security when channel noise cannot guarantee security and privacy.
	\item
	We formulate an integer nonlinear fractional optimization problem in the case of insufficient channel noise. For the special case where the model is with high dimension, the closed-form solution is obtained and useful insights are drawn. A BnB-based algorithm is proposed to solve this problem in the general case, which could achieve the optimal solution with low computational complexity.

	\item Finally, simulations are conducted to verify the effectiveness of the proposed algorithms and their superiority over conventional schemes.
\end{itemize}

\subsection{Organization}
The remainder of this paper is organized as follows. In Section \ref{Systemmodel}, we present the system model and introduce the procedure of FL, the definitions of DP and MSE-security. The details of SP-OTA-FL and CWPP mechanism are introduced in Section \ref{section3}, where we also conduct the privacy, security and convergence analysis of the proposed SP-OTA-FL. We propose device selection policies in Section \ref{section4}. The simulation results are shown in Section \ref{simulation} and we conclude the paper in Section \ref{conclusion}.

\section{System Model and preliminaries}~\label{Systemmodel}
In this section, the considered system model is presented. We introduce the basic concepts and procedure of FL in Section \ref{federatedlearning}. The definitions of DP and MSE-security used to measure the privacy leakage and security level of the system are introduced in Section \ref{differentialprivacy} and Section \ref{securityofsystem}, respectively. Typical notations used in this paper are summarized in Table I.

\begin{table*}[ht] 
	\caption{Summary of main notations}
	\centering
	\begin{tabular}{cc}
		\toprule
		Notation &  Description\\
		\midrule
		$\left\| \boldsymbol{x} \right\| _{\varsigma}$ & $\varsigma$-norm of vector $x$\\
		$\left| \mathcal{A} \right|$  &  Size of set $\mathcal{A}$ \\
		$\mathcal{N}$; $N$  & Set of all the devices; Size of $\mathcal{N}$ \\
		$T$ & Number of total training rounds\\
		$\mathcal{K} ^t$; $\mathcal{J} ^t$ & Set of uploaders in round $t$; Set of jammers in round $t$ \\
		$\mathcal{D} _n$; $D_n$ &  Dataset of device $n$; Size of $\mathcal{D} _n$\\
		$\mathcal{B} _n$; $B_n$ &  A batch of $\mathcal{D} _n$; Size of $\mathcal{B} _n$\\
		$L\left( \cdot \right)$; $L_n\left( \cdot \right)$; $l\left( \cdot;\cdot \right)$ & Global objective function; Local objective function of device $n$; Loss function\\
		$\boldsymbol{m }$; $d$ & Model parameter; Dimension of $m$ \\
		$\tau^t$ & Learning rate at round $t$\\
		$\boldsymbol{g}_{n}^{t}$; $\nabla L_n\left( \boldsymbol{m}_{n}^{t} \right)$; & Stochastic gradient of device $n$ in round $t$; Full gradient of device $n$ in round $t$\\	
		$\boldsymbol{e}_{n}^{t}$ & Jamming siganl (Gaussian artificial noise)\\
		$G$, $\vartheta$ & Upper bound of $\boldsymbol{g}_{n}^{t}$; Standard deviation of the stochastic gradient\\
		$\boldsymbol{x}_{n}^{t}$; $\boldsymbol{y}^t$; $\boldsymbol{z}^t$ & Input signal of device $n$ in round $t$; Received signal at BS; Received signal at Eve\\
		$\theta $ ;$\rho$ & Smoothness of $L_n\left( \cdot \right)$; Convexity of $L_n\left( \cdot \right)$\\
		$h_{n,B}^{t}$; $h_{n,E}^{t}$ & Channel gain between device $n$ and BS; Channel gain between device $n$ and eavesdropper\\
		$\boldsymbol{r}_{B}^{t}$; $\boldsymbol{r}_{E}^{t}$ & Channel noise at the BS; channel noise at Eve\\
		$\sigma _B$; $\sigma _E$ & Variance of receiver noise at the BS; Variance of receiver noise at eavesdropper\\	
		$P_n$ & Maximum transmission power of device $n$\\
		$\left( \epsilon,\zeta \right)$-DP; $\varDelta S$ & DP level; Sentivisity\\
		$\left( \mathcal{E} ,\phi \right)$-MSE-security & Security level of mechanism $\mathcal{E}$ \\
		\bottomrule
	\end{tabular}
\end{table*}
We consider an SP-OTA-FL system where $N$ edge devices, indexed by the set $\mathcal{N} =\left\{ 1,2,..., N \right\}$, collaboratively train a deep neural network (DNN) model with the help of a BS. The devices and BS communicate through a shared MAC where all devices transmit their gradients simultaneously using the same channel. As a result of the waveform-superposition property of MAC, the gradients are aggregated over the air. Specifically, the BS is assumed to be ``honest but curious" that may attempt to learn the personal information from the received gradients, which is regarded as a privacy threat. 
Additionally, the security threat is that there is an eavesdropper (Eve) in the system that tries to wiretap the gradients. 

\begin{figure*}[ht]
	\centering
	\subfigure[SP-OTA-FL with full device participation]	
	{  \begin{minipage}[t]{0.48\linewidth}
			\centering
			\includegraphics[scale=0.85]{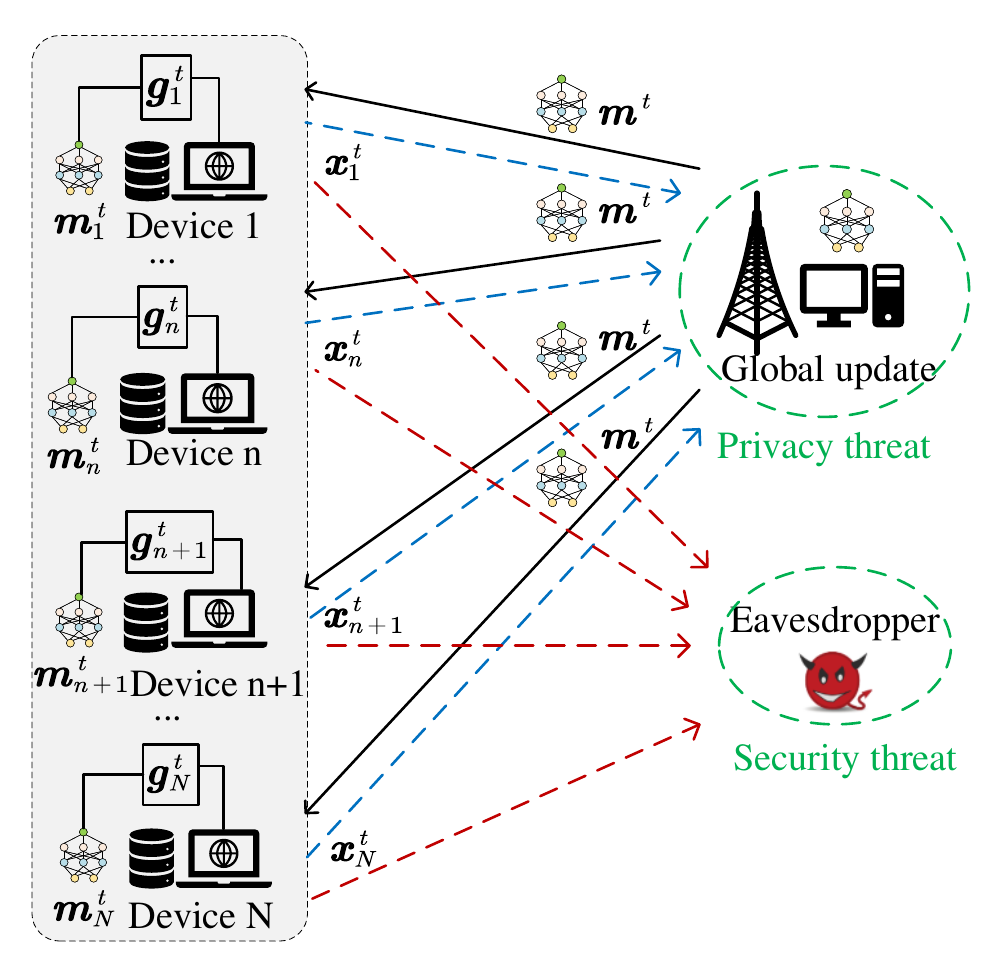}
		\end{minipage}	
	}
	{  \begin{minipage}[t]{0.48\linewidth}
		\centering
		\includegraphics[scale=0.85]{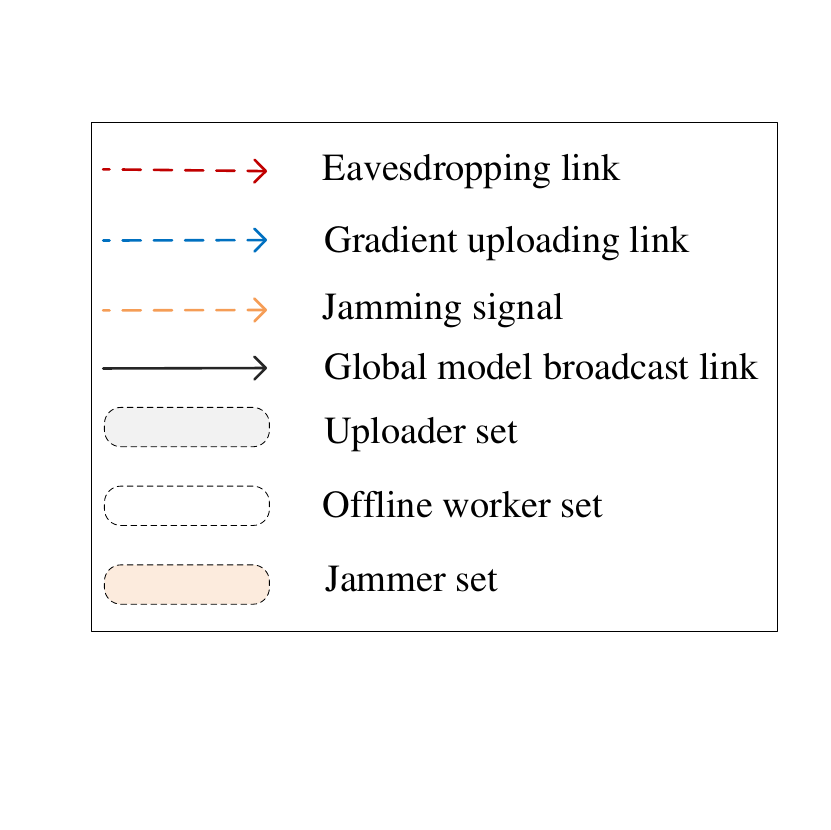}
	\end{minipage}	
	}
	\subfigure[SP-OTA-FL with partial device participation]	
	{  \begin{minipage}[t]{0.48\linewidth}
			\centering
			\includegraphics[scale=0.85]{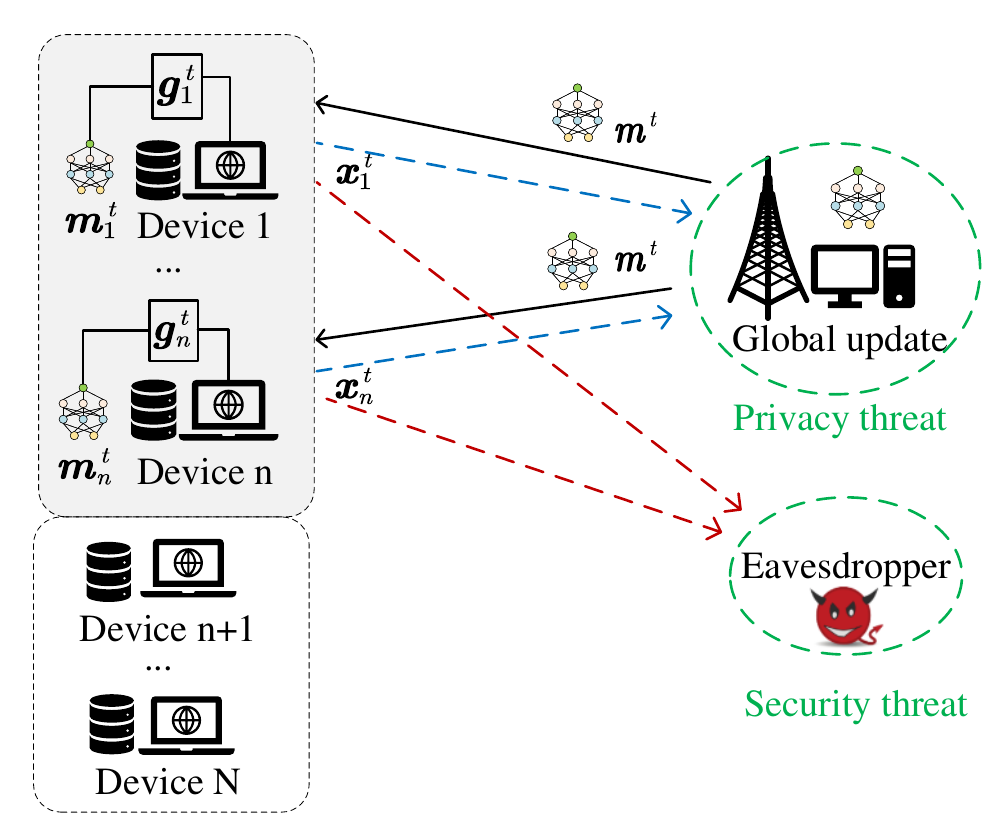}
		\end{minipage}	
	}
	\subfigure[Jamming-aided SP-OTA-FL]	
	{  \begin{minipage}[t]{0.48\linewidth}
			\centering
			\includegraphics[scale=0.85]{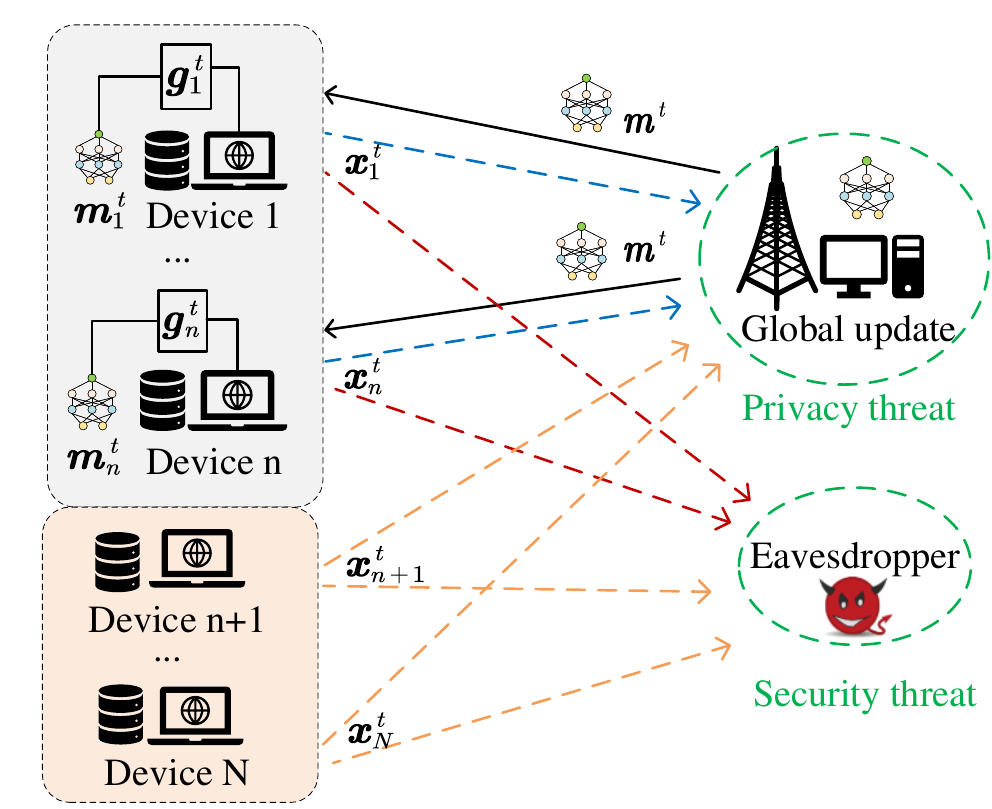}
		\end{minipage}	
	}
	
	\caption{The secure and private over-the-air federated learning.}
	\label{figsystemmodel}
\end{figure*}

To prevent the privacy leakage of user data and the security attack, in the considered system, we employ channel noise and artificial noise as protection in different cases. More specifically, three scenarios are considered as shown in Fig. \ref{figsystemmodel}. In the case that channel noise is sufficient for privacy and security protection, all the devices are selected to participate in the training process as shown in Fig. \ref{figsystemmodel} (a). Fig. \ref{figsystemmodel} (b) and Fig. \ref{figsystemmodel} (c) are proposed in the case of insufficient channel noise. In Fig. \ref{figsystemmodel} (b), we only select devices that the privacy and security can be guaranteed by channel noise as uploaders and others, referred to as offline workers, will be absent in the training. In the third case, part of the devices are selected as uploaders and others are selected as jammers to send Gaussian artificial noise for protecting privacy and security, as shown in Fig. \ref{figsystemmodel} (c).  The details of SP-OTA-FL and the three cases are described in Section \ref{section3} and Section \ref{section4}, respectively.



\subsection{Federated Learning}\label{federatedlearning}
	In the considered FL network, each device of index $n\in \mathcal{N}$ is assumed to have a local dataset $\mathcal{D} _n$ which contains $D_n$ pairs of training samples $\left( \boldsymbol{u},v \right)$ where $\boldsymbol{u}$ is the raw data and $v$ is the corresponding label. For simplicity, we assume that $D_1=\cdot \cdot \cdot =D_N$. 

Typically, the purpose of the FL task is to obtain the model
parameter that can minimize the loss function. Mathematically, the goal of the learning can be expressed as follows:
\begin{equation}\label{globallossfunction}
	\begin{aligned}
		\min_{\boldsymbol{m}} L\left( \boldsymbol{m} \right) =\frac{1}{N}\sum_{n=1}^N{L_n\left( \boldsymbol{m} \right)},
	\end{aligned}
\end{equation}
where $\boldsymbol{m }\in \mathbb{R} ^d$ is the model parameter to be optimized. More specifically, the objective function of device $n$ is defined as follows:
\begin{equation}\label{wg15}
	\begin{aligned}
				L_n\left( \boldsymbol{m} \right) =\frac{1}{D_n}\sum_{\left( \boldsymbol{u},v \right) \in \mathcal{D} _n}{l\left( \boldsymbol{m};\left( \boldsymbol{u},v \right) \right)},
	\end{aligned}
\end{equation}
where $l\left( \boldsymbol{m};\left( \boldsymbol{u},v \right) \right)$ is an empirical loss function defined by learning task, quantifying the loss of $\boldsymbol{m}$ at sample $\left( \boldsymbol{u},v \right)$. Some typical loss functions $l\left( \boldsymbol{m};\left( \boldsymbol{u},v \right) \right)$ applied in ML are listed in Table II. 

\begin{table*}[htbp] 
	\caption{Loss functions for popular machine learning models}
	\centering
	\begin{tabular}{cc}
		\toprule
		Model &  Loss function $l\left( \boldsymbol{m};\left( \boldsymbol{u},v \right) \right)$\\
		\midrule
		Linear regression & $\frac{1}{2}\left\| v_i-\boldsymbol{m}^T\boldsymbol{u}_i \right\| ^2$\\
		Squared-SVM & $\frac{\iota}{2}\left\| \boldsymbol{m} \right\| ^2+\frac{1}{2}\max \left\{ 0,1-v_i\boldsymbol{m}^T\boldsymbol{u}_i \right\} 
		$ where $\iota$ is a constant\\
		Neural network & Cross-entropy on cascaded linear and non-linear transform, see \cite{lecun2015deep} for details.\\
		\bottomrule
	\end{tabular}
\end{table*}

To solve the problem in (\ref{globallossfunction}), an iterative approach referred to as gradient descent (GD) can be applied. However, it may be impractical to perform GD over the whole local dataset because of the considerably massive data samples in reality. Alternatively, stochastic gradient descent (SGD) as one of the practical solutions is more widely used in FL as it computes the gradient over a batch of data samples, randomly chosen from the local dataset, as an approximation of the full gradient (obtained based on the whole dataset). The main procedure of basic SGD applied in FL is given as follows:
\begin{itemize}
	\item \textbf{Step 1}: \emph{Parameters broadcasting}: At the beginning of round $t$, the BS first selects part of the devices to participate in the current training round and broadcasts the latest global model parameter $\boldsymbol{m}^{t}$ to these devices.

	\item \textbf{Step 2}: \emph{Local training}: (1) Each participant performs the initialization of the local model by setting the received global model parameter as the local model parameter, i.e., $\boldsymbol{m}_{n}^{t} = \boldsymbol{m}^{t}, \forall n$. (2) Each device randomly selects a batch of data $\mathcal{B} _n$ of size $B_n$ from $\mathcal{D} _n$ and computes the stochastic gradient based on $\mathcal{B} _n$. More specifically, the stochastic gradient is given by
	\begin{equation}\label{wg15}
		\begin{aligned}
			\boldsymbol{g}_{n}^{t}\triangleq \nabla L_n\left( \boldsymbol{m}_{n}^{t};\mathcal{B} _n \right) =\frac{1}{B_n}\sum_{\left( \boldsymbol{u},v \right) \in \mathcal{B} _n}{\nabla l\left( \boldsymbol{m}_{n}^{t};\left( \boldsymbol{u},v \right) \right)}.
		\end{aligned}
	\end{equation}
	By contrast, the full gradient in GD is given by
	\begin{equation}\label{wg15}
		\begin{aligned}
			\nabla L_n\left( \boldsymbol{m}_{n}^{t} \right) =\frac{1}{D_n}\sum_{\left( \boldsymbol{u},v \right) \in \mathcal{D} _n}{\nabla l\left( \boldsymbol{m}_{n}^{t};\left( \boldsymbol{u},v \right) \right)}.
		\end{aligned}
	\end{equation}

	\item \textbf{Step 3}: \emph{Gradients aggregation}: (1) Devices send the obtained gradients to the BS.
	(2) Upon receiving all the gradients from the participants, the BS makes aggregation of the received gradients as follows:
	\begin{equation}\label{wg15}
		\begin{aligned}
			\boldsymbol{g}^t=\sum_{n\in \mathcal{K} ^t}{w_{n}^{t}\boldsymbol{g}_{n}^{t}},
		\end{aligned}
	\end{equation}
	where $w_{n}^{t}$ is the weight of gradient from device $n$ in round $t$ and satisfies $\sum_{n\in \mathcal{K} ^t}{w_{n}^{t}=1}$. In most of existing studies, $w_{n}^{t}=\frac{B_n}{B}$ where $B=\sum_{n\in \mathcal{K} ^t}{B_n}$. 
	\item \textbf{Step 4}: \emph{Model update}: The BS performs global model update as follows:
	\begin{equation}\label{globalupdate}
		\begin{aligned}
			\boldsymbol{m}^{t+1}=\boldsymbol{m}^t-\tau ^t \boldsymbol{g}^t,
		\end{aligned}
	\end{equation}
	where $\tau ^t$ is the learning rate (also termed as step size in SGD). 
\end{itemize}
The above iteration steps are repeated until a certain training termination condition is met.


In order to formally quantify the privacy leakage and the security level of the system, we introduce the DP and MES-security concepts in the following.
\subsection{Differential Privacy}\label{differentialprivacy}
DP \cite{dwork2014algorithmic} is defined on the conception of the adjacent dataset, which guarantees the probability that any two adjacent datesets output the same result is less than a constant with the help of adding random noise. More specifically, DP quantifies information leakage in FL by measuring the sensitivity of the disclosed statistics (i.e., the gradients) to the change of a single data point in the input dataset. The basic definition of $\left( \epsilon ,\zeta \right)$-DP is given as follows.
\newtheorem{definition}{Definition}
\begin{definition}$\left( \epsilon ,\zeta \right)$-DP \cite{dwork2014algorithmic}:
	A randomized mechanism $\mathcal{O}$ guarantees $\left( \epsilon ,\zeta \right)$-DP if for two adjacent datasets $\mathcal{D} ,\mathcal{D} '$ differing in one sample, and measurable output space $\mathcal{Q} $ of $\mathcal{O}$, it satisfies,
	\begin{equation}\label{wg11}
		\begin{aligned}
			\mathrm{Pr}\left[ \mathcal{O} \left( \mathcal{D} \right) \in \mathcal{Q} \right] \leqslant e^{\epsilon}\mathrm{Pr}\left[ \mathcal{O} \left( \mathcal{D} ' \right) \in \mathcal{Q} \right] +\zeta.
		\end{aligned}
	\end{equation} 
\end{definition}
The additive term $\zeta$ allows for breaching $\epsilon$-DP with the probability $\zeta$ while $\epsilon$ denotes the protection level and a smaller $\epsilon$ means a higher privacy preservation level. Specifically, the Gaussian DP mechanism which guarantees privacy by adding artificial Gaussian noise is introduced as follows.
\begin{definition}Gaussian mechanism \cite{dwork2014algorithmic}:
	A mechanism $\mathcal{O}$ is called as a Gaussian mechanism, which alters the output of another algorithm $\mathcal{L} :\mathcal{D} \rightarrow \mathcal{Q}$ by adding Gaussian noise, i.e., 
	\begin{equation}\label{wg11}
		\begin{aligned}
			\mathcal{O} \left( \mathcal{D} \right) =\mathcal{L} \left( \mathcal{D} \right) +\mathcal{N} \left( 0,\sigma ^2\mathbf{I}_d \right).
		\end{aligned}
	\end{equation} 
	Gaussian mechanism $\mathcal{O}$ guarantees $\left(\epsilon ,\zeta \right)$-DP with $\epsilon =\frac{\varDelta S}{\sigma}\sqrt{2\ln \left( \frac{1.25}{\zeta} \right)}$ where $\varDelta S\triangleq \underset{\mathcal{D} ,\mathcal{D} '}{\max}\left\| \mathcal{L} \left( \mathcal{D} \right) -\mathcal{L} \left( \mathcal{D} ' \right) \right\| _2$ is the sensitivity of the algorithm $\mathcal{L}$  quantifying the sensitivity of the algorithm $\mathcal{L}$ to the change of a single data point. 
\end{definition}
According to the Gaussian mechanism described above, privacy leakage depends both on the sensitivity of the algorithm $\mathcal{L}$ and on the power of the added Gaussian noise.

\subsection{MSE Security}\label{securityofsystem}
In this work, the gradients are transmitted in an analog way and aggregated via AirComp. In accordance with \cite{frey2020towards}, the way to improve the security of an analog AirComp setting, is to employ noise as jamming to degrade the eavesdropper's SNR and thus prevent it from recovering a low-noise estimate of the transmitted message. MSE-security has been proposed in \cite{frey2020towards} to measure the security of analog messages and is introduced as follows.


\begin{definition}\label{difinition3} $\left( \mathcal{E} ,\phi \right)$-MSE-security \cite{frey2020towards}:
	A uniform distributed mechanism $\mathcal{E} : \mathcal{G} \rightarrow \mathcal{Y}$, where $\mathcal{Y}$ is a measureable and bounded output space, guarantees $\left( \mathcal{E} ,\phi \right)$-MSE-security if under a uniform distribution of 
	$\mathcal{E} \left( \left\{ \boldsymbol{g}_{n}^{t} \right\} _{n\in \mathcal{K} ^t} \right)$, for any Eve's estimator $e:\mathcal{Z} \rightarrow \mathcal{Y} $, there is a real number $\phi \geqslant 0$ satisfies,
	\begin{equation}\label{wg11}
		\begin{aligned}
			\mathbb{E} \left[ \left( e\left( \boldsymbol{z}^t \right) -\mathcal{E} \left( \left\{ \boldsymbol{g}_{n}^{t} \right\} _{n\in \mathcal{K} ^t} \right) \right) ^2 \right] \geqslant \phi.
		\end{aligned}
	\end{equation} 
\end{definition}
In statistical terms, a scheme guarantees $\left( \mathcal{E},\phi \right)$-MSE-security means that all estimators that the eavesdropper can apply have MSE at least $\phi$.


\section{SP-OTA-FL Framework}~\label{section3}

In the proposed framework, we employ Gaussian noise to improve the security and privacy inspired by DP and the AirComp security \cite{frey2020towards}. Given the condition that channel noise may be insufficient for all the devices to participate in training with security and privacy protection, we assume that the BS selects some of the devices as uploaders in each training round $t$, denoted by $\mathcal{K}^t \subseteq \mathcal{N}$, to participate in the training, and selects some devices as jammers in some cases, denoted by $\mathcal{J}^t  \subseteq \mathcal{N} \setminus \mathcal{K}^t$, to send Gaussian artificial noise to enhance system security and protect privacy. Particularly, $\mathcal{J} ^t$ is empty in the cases that there is no device selected as jammers. 

Assume that the upper bound of $\left\| \boldsymbol{g}_{n}^{t} \right\| _2
$ is $G$. Given the selected uploader set $\mathcal{K} ^t$ and the jammer set $\mathcal{J} ^t$, we next present the details of SP-OTA-FL. 
The signal from device $n$ is given by
\begin{equation}\label{wg15}
	\begin{aligned}	\	
		\boldsymbol{x}_{n}^{t}=\left\{ \begin{array}{c}
			\frac{\sqrt{P_n}}{G}\boldsymbol{g}_{n}^{t},n\in \mathcal{K} ^t\\
			\sqrt{\frac{P_n}{d}}\boldsymbol{e}_{n}^{t},n\in \mathcal{J} ^t\\
		\end{array} \right.   
	\end{aligned}
\end{equation}
where $P_n$ is the maximum transmission power of device $n$ and $\boldsymbol{e}_{n}^{t}\sim \mathcal{N} \left( 0,\mathbf{I}_d \right)$ is the artificial noise sent from jammer $n$ in round $t$. The $h_{n,B}^{t}\in \mathbb{R} ^+$, $h_{n,E}^{t}\in \mathbb{R} ^+$ are the channel gain coefficients between device $n$ and the BS and that between device $n$ and the eavesdropper, respectively. We assume real channel gain coefficients for simplicity \cite{hasirciouglu2021private}. The coefficients are independent across devices and training rounds but remain constant within one round. Consequently, the received signals at the BS and eavesdropper in round $t$ are given as follows:
\begin{equation}\label{wg15}
	\begin{aligned}
		\boldsymbol{y}^t=&\sum_{n=1}^N{h_{n,B}^{t}\boldsymbol{x}_{n}^{t}}+\boldsymbol{r}_{B}^{t}
		\\
		=&\sum_{n\in \mathcal{K} ^t}{\frac{h_{n,B}^{t}\sqrt{P_n}}{G}\boldsymbol{g}_{n}^{t}}+\sum_{n\in \mathcal{J} ^t}{h_{n,B}^{t}\sqrt{\frac{P_n}{d}}\boldsymbol{e}_{n}^{t}}+\boldsymbol{r}_{B}^{t},
	\end{aligned}
\end{equation}
\begin{equation}\label{receivedsignalEve}
	\begin{aligned}
		\boldsymbol{z}^t=&\sum_{n=1}^N{h_{n,E}^{t}\boldsymbol{x}_{n}^{t}}+\boldsymbol{r}_{E}^{t}
		\\
		=&\sum_{n\in \mathcal{K} ^t}{\frac{h_{n,E}^{t}\sqrt{P_n}}{G}\boldsymbol{g}_{n}^{t}}+\sum_{n\in \mathcal{J} ^t}{h_{n,E}^{t}\sqrt{\frac{P_n}{d}}\boldsymbol{e}_{n}^{t}}+\boldsymbol{r}_{E}^{t},	
	\end{aligned}
\end{equation}
where $\boldsymbol{r}_{B}^{t}\sim \mathcal{N} \left( 0,\sigma _B\mathbf{I}_d \right) $ and $\boldsymbol{r}_{E}^{t}\sim \mathcal{N} \left( 0,\sigma _E\mathbf{I}_d \right)$ are the received noise at the BS and eavesdropper, respectively. For ease of presentation, we use
\begin{equation}\label{wg13}
	\begin{aligned}
		\boldsymbol{r}_{B,Tot}^{t}=\sum_{n\in \mathcal{J} ^t}{h_{n,B}^{t}\sqrt{\frac{P_n}{d}}\boldsymbol{e}_{n}^{t}}+\boldsymbol{r}_{B}^{t},
	\end{aligned}
\end{equation}	
\begin{equation}\label{wg14}
	\begin{aligned}
		\boldsymbol{r}_{E,Tot}^{t}=\sum_{n\in \mathcal{J} ^t}{h_{n,E}^{t}\sqrt{\frac{P_n}{d}}\boldsymbol{e}_{n}^{t}}+\boldsymbol{r}_{E}^{t},
	\end{aligned}
\end{equation}
to denote the aggregated noise at the BS and eavesdropper, respectively.

\subsection{CWPP Mechanism}
Similar to \cite{yan2022performance}, we propose a channel-weighted aggregation scheme to alleviate the negative impact of noise on the utility of aggregated gradient. By employing channel-weighted aggregation, the gradient does not need to be aligned during the transmission and therefore, can avoid the issue that the SNR of the system will be limited by the device with the worst channel quality.  

In order to recover an estimate of averaging gradient from the aggregated gradients, the BS performs post-processing by,
\begin{equation}\label{channel-weighted-1}
	\begin{aligned}
		\boldsymbol{\tilde{g}}^t=&\frac{G}{H^t}\boldsymbol{y}_{B}^{t}
		\\
		=&\frac{G}{H^t}\left( \sum_{n\in \mathcal{K} ^t}{\frac{h_{n,B}^{t}\sqrt{P_n}}{G}\boldsymbol{g}_{n}^{t}}+\boldsymbol{r}_{B,Tot}^{t} \right),
	\end{aligned}
\end{equation}
where $H^t=\sum_{n\in \mathcal{K} ^t}{h_{n,B}^{t}\sqrt{P_n}}$. More specifically, an insight into the CWPP scheme can be given by
\begin{equation}\label{channel-weighted-2}
 	\begin{aligned}
	 		\boldsymbol{\tilde{g}}^t=&\frac{1}{H^t}\left( \sum_{n\in \mathcal{K} ^t}{h_{n,B}^{t}\sqrt{P_n}\boldsymbol{g}_{n}^{t}}+G\boldsymbol{r}_{B,Tot}^{t} \right) 
	 		\\
	 		=&\sum_{n\in \mathcal{K} ^t}{\frac{h_{n,B}^{t}\sqrt{P_n}}{H^t}\left( \boldsymbol{g}_{n}^{t}+\frac{G\boldsymbol{r}_{B,Tot}^{t}}{h_{n,B}^{t}\sqrt{P_n}} \right)},
 	\end{aligned}
\end{equation}
from which one can find that, by performing the CWPP scheme, the gradient from the device with poor channel quality is assigned a smaller weight in the aggregation, thereby, mitigating the negative impact of noise on the learning process. 
Additionally, different from the conventional gradient-aligned OTA-FL \cite{seif2020wireless}, the power allocation in the CWPP scheme does not force gradient alignment. As a result, the overall SNR of the system will not be limited by the device with the worst channel condition.

\subsection{Privacy, Security and Convergence Analysis}

\subsubsection{Assumptions}
For analysis purposes, we provide the following assumptions first.
\newtheorem{assumption}{Assumption}
\begin{assumption}\label{gradientassumption}
	The assumptions on gradients:
	
	(1) Assume that the stochastic gradient is an unbiased estimate of the full gradient. 
	\begin{equation}\label{wg17}
		\begin{aligned}
			\mathbb{E} \left[ \boldsymbol{g}_{n}^{t} \right] =\nabla L_n\left( \boldsymbol{m}_{n}^{t} \right).
		\end{aligned}
	\end{equation}

(2) The variance of stochastic gradients at each device is bounded:
	\begin{equation}\label{wg18}
		\begin{aligned}
			\mathbb{E} \left[ \left\| \boldsymbol{g}_{n}^{t}-\nabla L_n\left( \boldsymbol{m}_{n}^{t} \right) \right\| _{2}^{2} \right] \leqslant \vartheta ^2.
		\end{aligned}
	\end{equation}

(3) The expected squared norm of stochastic gradients is bounded:
\begin{equation}\label{wg19}	
	\begin{aligned}
		\mathbb{E} \left[ \left\| \boldsymbol{g}_{n}^{t} \right\| _2 \right] \leqslant G.
	\end{aligned}
	\end{equation}
\end{assumption}

\begin{assumption}\label{assumption1}
	For each $n$, $L_n\left( \cdot \right) \,\,$ is $\theta$-smooth, i.e., for all $\boldsymbol{\iota }'$ and $\boldsymbol{\iota }$, one has
	\begin{equation}\label{wg8}
		\begin{aligned}
			L_n\left( \boldsymbol{\iota }' \right) -L_n\left( \boldsymbol{\iota } \right) \leqslant \left( \boldsymbol{\iota }'-\boldsymbol{\iota } \right) ^{\mathrm{T}}\nabla L_n\left( \boldsymbol{\iota } \right) +\frac{\theta}{2}\left\| \boldsymbol{\iota }'-\boldsymbol{\iota } \right\| _{2}^{2}.
		\end{aligned}
	\end{equation}
\end{assumption}

\begin{assumption}\label{assumption2}
	For each $n$, $L_n\left( \cdot \right) \,\,$ is $\rho$-strongly convex, i.e., for all $\boldsymbol{\iota }'$ and $\boldsymbol{\iota }$, one has
	\begin{equation}\label{wg9}
		\begin{aligned}
			L_n\left( \boldsymbol{\iota }' \right) -L_n\left( \boldsymbol{\iota } \right) \geqslant \left( \boldsymbol{\iota }'-\boldsymbol{\iota } \right) ^{\mathrm{T}}\nabla L_n\left( \boldsymbol{\iota } \right) +\frac{\rho}{2}\left\| \boldsymbol{\iota }'-\boldsymbol{\iota } \right\| _{2}^{2}.
		\end{aligned}
	\end{equation}
\end{assumption}

\begin{lemma}\label{lemma1}
	Assume that Assumption 2 and Assumption 3 hold. $L\left( \cdot \right)$ is $\theta$-smooth and $\rho$-strongly convex.
\end{lemma}

\textit{Proof:} : Please refer to Appendix \ref{proofofsmoothness}. $\hfill \blacksquare$


\subsubsection{Privacy analysis}
We here present the privacy analysis for SP-OTA-FL.
Following (\ref{wg13}), the variance of the aggregated noise at the BS is given by
\begin{equation}\label{wg15}
	\begin{aligned}
		\sigma _{B,Tot}^{t}=\sum_{n\in \mathcal{J} ^t}{h_{n,B}^{t}\sqrt{\frac{P_n}{d}}}+\sigma _B.
	\end{aligned}
\end{equation}

\begin{lemma}\label{lemma2}
	Assume that Assumption 1 holds. SP-OTA-FL guarantees $\left( \epsilon _{n}^{t},\zeta \right)$-DP of uploader $n$ in round $t$ when the following condition is satisfied,
	\begin{equation}\label{11111}
		\begin{aligned}
		  \epsilon _{n}^{t}=\frac{\varDelta S_{n}^{t}}{\sqrt{\sigma _{B,Tot}^{t}}}\sqrt{2\ln \left( \frac{1.25}{\zeta} \right)},
		\end{aligned}
	\end{equation}
where $\varDelta S_{n}^{t}=2h_{n,B}^{t}\sqrt{P_n}$.
\end{lemma} 

\textit{Proof:} : Please refer to Appendix \ref{proofofprivacyanalysis}. $\hfill \blacksquare$

Lemma \ref{lemma2} reveals an important insight that devices with better channel quality are more prone to privacy disclosure. Therefore, for reducing the privacy leakage in the system, one can either increase the power of noise or select devices with smaller channel condition coefficient $h_{n, B}^{t}\sqrt{P_n}$ to participate in training.

\begin{remark}
	Note that when the ``$=$" in $\varDelta S_{n}^{t}=2h_{n,B}^{t}\sqrt{P_n}$ is replaced by ``$\leqslant$", it indicates a stronger privacy protection so it still satisfies $\left( \epsilon _{n}^{t},\zeta \right)$-DP.
\end{remark}

\subsubsection{Security analysis}
We here present the security analysis for SP-OTA-FL. Assume that the goal of eavesdropper is to recover an averaging estimate of the gradients, denoted by $\boldsymbol{g}_{ave}^{t}=\frac{1}{\left| \mathcal{K} ^t \right|}\sum_{n\in \mathcal{K} ^t}{\boldsymbol{g}_{n}^{t}}$, which can be used for global model update and further exploring the sensitive information of each device. We define the aggregation mechanism in (\ref{receivedsignalEve}) as $\mathcal{E} ^t:\left( \boldsymbol{g}_{n}^{t} \right) _{n\in \mathcal{K} ^t}\rightarrow \boldsymbol{z}^t\in \mathcal{Z}$, then we have the following result.  

\begin{lemma} \label{securitylemma}
Assume that the elements of $\boldsymbol{g}_{n}^{t}$ are distributed uniformly in [a, b]. The aggregation mechanism $\mathcal{E} ^t:\left( \boldsymbol{g}_{n}^{t} \right) _{n\in \mathcal{K} ^t}\rightarrow \boldsymbol{z}^t\in \mathcal{Z}$ guarteens $\left( \mathcal{E} ^t,\gamma _{E}^{t}\varXi \left( \frac{b-a}{\sqrt{\gamma _{E}^{t}}} \right) \right)$-MSE-security in training round $t$. Specifically, 
\begin{equation}\label{securitylemmawg1}
	\begin{aligned}
			\gamma _{E}^{t}=\frac{G^2}{\left| \mathcal{K} ^t \right|\left( \varLambda ^t \right) ^2}\left( \sum_{n\in \mathcal{J} ^t}{\frac{\left( h_{n,E}^{t} \right) ^2P_n}{d}}+\sigma _E\,\, \right),
	\end{aligned}
\end{equation}
where $\varLambda ^t=\underset{n\in \mathcal{K} ^t}{\max}\left\{ h_{n,B}^{t}\sqrt{P_n} \right\}  
$, and
\begin{equation}\label{securitylemmawg2}
	\begin{aligned}
		\varXi \left( t \right) =\int_0^t{\int_{-\infty}^{+\infty}{\left( v+\frac{\varphi _N\left( -v \right) -\varphi _N\left( t-v \right)}{\varPhi _N\left( t-v \right) -\varPhi _N\left( -v \right)}-u \right)}}^2
		\cdot \frac{1}{t}\varphi _N\left( u-v \right) dvdu,
	\end{aligned}
\end{equation}
	where $\varphi _N\left( \cdot \right)$ and $\varPhi _N\left( \cdot \right)$ denote the probability density function and the cumulative distribution function of the standard normal distribution. 
\end{lemma} 
\textit{Proof:} Please refer to Appendix \ref{proofofsecuritylemma}.$\hfill \blacksquare$

According  to Definition \ref{difinition3}, $\left( \mathcal{E} ^t,\gamma _{E}^{t}\varXi \left( \frac{b-a}{\sqrt{\gamma _{E}^{t}}} \right) \right)$-MSE-security means that the gradient estimates recovered from $\boldsymbol{z}^t$ are with the MSE at least $\gamma _{E}^{t}\varXi \left( \frac{b-a}{\sqrt{\gamma _{E}^{t}}} \right)$. It has been validated that $\gamma _{E}^{t}\varXi \left( \frac{b-a}{\sqrt{\gamma _{E}^{t}}} \right)$ increases with $\gamma _{E}^{t}$  in \cite{frey2020towards}. Therefore, a bigger $\gamma _{E}^{t}$ means higher system security and we use $\gamma _{E}^{t}$ to indicate the security level of the system which is referred to as the security coefficient. Similar to the privacy analysis, (\ref{securitylemmawg1}) proves that one way to secure the FL process is to increase the aggregated noise at Eve or to select devices with relatively poor channel conditions to participate in the training to make a smaller $\varLambda^t$. 
\subsubsection{Convergence analysis}\label{convergenceanalysis}
For ease of presentation, we denote $p_{n,B}^{t}=h_{n,B}^{t}\sqrt{P_n}$ and $p_{n,E}^{t}=h_{n,E}^{t}\sqrt{P_n}$. For analytical tractability, we define 
\begin{equation}\label{noiselessSGD}
	\begin{aligned}
		\boldsymbol{\hat{g}}^t=&\sum_{n\in \mathcal{K} ^t}{\frac{p_{n,B}^{t}}{\sum_{n\in \mathcal{K} ^t}{p_{n,B}^{t}}}\boldsymbol{g}_{n}^{t}},
	\end{aligned}
\end{equation}
\begin{equation}\label{wg26}
	\begin{aligned}
		\boldsymbol{\bar{g}}^t=&\sum_{n\in \mathcal{K} ^t}{\frac{p_{n,B}^{t}}{\sum_{n\in \mathcal{K} ^t}{p_{n,B}^{t}}}\nabla L_n\left( \boldsymbol{m}_{n}^{t} \right)},
	\end{aligned}
\end{equation}
to denote the noise-free aggregated stochastic gradient and full gradient, respectively.
Then, it thus follows (\ref{globalupdate}), (\ref{wg13}), (\ref{channel-weighted-1}) and (\ref{noiselessSGD}) that the update of the global model performed by the BS can be given by
\begin{equation}\label{globalmodeupdate}
	\begin{aligned}
		\boldsymbol{m}^{t+1}=\boldsymbol{m}^t-\tau ^t\left( \boldsymbol{\hat{g}}^t+\frac{G}{\sum_{n\in \mathcal{K} ^t}{p_{n,B}^{t}}}\boldsymbol{r}_{B,Tot}^{t} \right) ,
	\end{aligned}
\end{equation}
where $\boldsymbol{r}_{B,Tot}^{t}=\sum_{n\in \mathcal{J} ^t}{\frac{p_{n,B}^{t}}{\sqrt{d}}\boldsymbol{e}_{n}^{t}}+\boldsymbol{r}_{B}^{t}$. Then, we have the following results.
\begin{lemma}\label{lemma4}
	Assume that Assumption 1 holds. The noise-free aggregated stochastic gradient is an unbiased estimate of the noise-free aggregated full gradient, i.e.,
	\begin{equation}\label{optimalitygap}
		\begin{aligned}
			\mathbb{E} \left[ \boldsymbol{\hat{g}}^t-\boldsymbol{\bar{g}}^t \right] =0.
		\end{aligned}
	\end{equation}
The variance of the noise-free aggregated stochastic gradient is bounded:
	\begin{equation}\label{optimalitygap}
		\begin{aligned}
			\mathbb{E} \left[ \left\| \boldsymbol{\hat{g}}^t-\boldsymbol{\bar{g}}^t \right\| _{2}^{2} \right] \leqslant \vartheta ^2.
		\end{aligned}
	\end{equation}
\end{lemma} 
\textit{Proof:} Please refer to Appendix \ref{proofoflemma4}.$\hfill \blacksquare$
\begin{lemma}\cite{yan2022performance}\label{lemma5}
	Assume that Assumption 1 holds and  $\boldsymbol{m}^*=\left[ m_{1}^{*},\cdots ,m_{d}^{*} \right] $, $\boldsymbol{m}_{n}^{*}=\left[ m_{n,1}^{*},\cdots ,m_{n,d}^{*} \right] $ are the globally optimal model and the locally optimal model of device $n$, respectively. Then, for each device $n$, the upper bound of the gap between $L_n\left( \boldsymbol{m}^* \right)$ and $L_n\left( \boldsymbol{m}_{n}^{*} \right)$ is given by
	\begin{equation}\label{wg122}
		\begin{aligned}
			L_n\left( \boldsymbol{m}^* \right) -L_n\left( \boldsymbol{m}_{n}^{*} \right) \leqslant \varGamma,
		\end{aligned}
	\end{equation}
	where $\varGamma =\underset{n}{\max}\left\{ \frac{\theta d}{2}\left( \underset{i}{\max}\left\{ \left| m_{i}^{*}-m_{n,i}^{*} \right| \right\} \right) ^2 \right\} $.
\end{lemma} 

Furthermore, if the data is IID, $\varGamma$ goes to zero as the number of samples approaches infinity \cite{Li2020On}. 
							
\begin{theorem}\label{lemmaconvergenceanalysis}
	Assume that Assumption 1 to Assumption 3 hold and let $\frac{1}{\varrho}\leqslant \tau ^t\leqslant \frac{1}{\theta}$ with $\varrho$ been a constant. The bound of the gap between model $\boldsymbol{m}^{t+1}$ and the optimal model $\boldsymbol{m}^*$ is given by
	\begin{equation}\label{optimalitygap}
		\begin{aligned}
			\mathbb{E} \left[ \left\| \boldsymbol{m}^{t+1}-\boldsymbol{m}^* \right\| _{2}^{2} \right] \leqslant \left( 1-\rho \tau ^t \right) \mathbb{E} \left[ \left\| \boldsymbol{m}^t-\boldsymbol{m}^* \right\| _{2}^{2} \right] +\left( \tau ^t \right) ^2\left( 2\varrho \varGamma +\vartheta ^2+G^2\varPsi ^t \right) ,
		\end{aligned}
	\end{equation}
where 
\begin{equation}\label{noiseimpact}
	\begin{aligned}
		\varPsi ^t=\frac{N\sum_{n\in \mathcal{J} ^t}{\left( p_{n,B}^{t} \right) ^2}+d\sigma _B}{\left( \sum_{n\in \mathcal{K} ^t}{p_{n,B}^{t}} \right) ^2},
	\end{aligned}
\end{equation}
which characterizes the impact of the device schedule in training round $t$.
The expectation is with respect to the stochastic gradient function and the randomness of Gaussian noise.
\end{theorem} 
\textit{Proof:} Please refer to Appendix \ref{proofofoptimalitygap}.$\hfill \blacksquare$

In (\ref{noiseimpact}), the impact of noise and the device selection on learning performance has been theoretically illustrated. According to Lemma \ref{lemma2} and Lemma \ref{securitylemma}, noise and low power of gradient transmission contribute to the security and privacy protection, however, it has a negative impact on the learning process according to Theorem \ref{lemmaconvergenceanalysis}. The scale of the noise and the power of the uploaded gradients depend on the scheduling of the devices. Therefore, an appropriate device selection decision is significant for SP-OTA-FL.


	
\begin{corollary} \label{corollaryoptimalitygap}
		Assume that Assumption 1 to Assumption 3 hold. Let $\tau ^t=\frac{2}{\rho t+2\theta}$ and $\varrho =\frac{\rho T+2\theta}{2}$. When the training process terminates after $T$ rounds and $\boldsymbol{m}^T$ is returned as the final solution, the bound of the optimality gap can be given by
		\begin{equation}\label{goptimalitygap}
			\begin{aligned}
				\mathbb{E} \left[ L\left( \boldsymbol{m}^T \right) \right] -L^*\leqslant \frac{\theta}{\rho T+2\theta}\left[ \frac{2}{\rho}\left( \vartheta ^2+G^2\underset{t}{\max}\left\{ \varPsi ^t \right\} \right) \right] +\frac{2\varGamma \theta}{\rho}.
			\end{aligned}
		\end{equation}
\end{corollary}
\textit{Proof:} Please refer to Appendix \ref{proofofcorollary1}.$\hfill \blacksquare$

From (\ref{goptimalitygap}), one can find that the first term on the right hand side decreases with $T$, and will go to zero when $T$ approaches infinity, which implies that the proposed SP-OTA-FL can converge with the rate of $\mathcal{O} \left( \frac{1}{T} \right)$. In particular, if the data is IID, then the optimality gap goes to zero as the number of samples and the number of training rounds $T$ grow.

\section{Device Selection for SP-OTA-FL}~\label{section4}
In this section, we propose device scheduling strategies based on the above analytical results. For simplicity, we consider that all the devices have the same privacy constraint $\epsilon$ in all the training rounds and the $\varUpsilon$ is the coefficient in terms of security requirement. We take one round as an example to analyze the device selection process and therefore omit the index $t$ of the training round in the rest of the paper. We also define $\kappa =\sqrt{2\ln \left( \frac{1.25}{\zeta} \right)}$, $p_M=\underset{n}{\max}\left\{ p_{n,B} \right\} $ and $p_m=\underset{n}{\min}\left\{ p_{n,B} \right\}$ for ease of presentation. Then, we have the following analysis. 

Following Lemma \ref{lemma2} and Lemma \ref{securitylemma}, devices with poor channel conditions, i.e., smaller $p_{n,B}$ and $p_{n,E}$, have less risk at privacy leakage and security attack. Therefore, if the channel condition coefficients of all the devices in this system are lower than a certain critical point, the channel noise at the BS and Eve are enough to prevent privacy leakage and security attack. Specifically, we can obtain the critical point $\hat{p}=\min \left\{ \frac{\epsilon \sqrt{\sigma _B}}{2\kappa},\frac{G\sqrt{\sigma _E}}{N\sqrt{\varUpsilon}} \right\} $ by solving $\frac{2\hat{p}\kappa}{\sqrt{\sigma _B}}\leqslant \epsilon$ and $\frac{G\sqrt{\sigma _E}}{N\hat{p}}\geqslant \sqrt{\varUpsilon}$ (according to Lemma \ref{lemma2} and Lemma \ref{securitylemma}) where only the channel noise is considered for guaranteeing privacy and security. We replaced the $\left| \mathcal{K} ^t \right|$ in (\ref{securitylemmawg1}) with $N$ for simplicity, which offers a stronger security guarantee. Then, we propose the following policies in different cases of channel noise.

\subsubsection{Channel noise is sufficient for protecting privacy and security with all device participation} In the case of $p_M \leqslant \hat{p}$, the received noise at the BS and Eve is sufficient for the device with the best channel condition to participate in training while satisfying the privacy and security constraints. Therefore, all the devices can be selected as uploaders to participate in training in this round as shown in Fig. \ref{figsystemmodel} (a) in Section \ref{Systemmodel}.

\subsubsection{Channel noise is sufficient for protecting privacy and security with partial device participation}

In the case of $p_m\leqslant \hat{p}\leqslant p_M$, if no device is selected as a jammer that sends Gaussian artificial noise to increase the power of the aggregated noise, the received noise at the BS and Eve can only provide qualified privacy and security protection when the devices satisfying $p_n\leqslant \hat{p}$ are selected to participate in the training process. In such cases, there are two approaches to device scheduling.
\begin{itemize}
	\item Policy-1: Select those devices with $p_n\leqslant \hat{p}$ as uploaders to participate in the current training round, and other devices will be absent in this training round, as shown in Fig. \ref{figsystemmodel} (b) in Section \ref{Systemmodel}.
	
	\item Policy-2: Select some devices as uploaders and others as jammers, as shown in Fig. \ref{figsystemmodel} (c) in Section \ref{Systemmodel}, via solving optimization problems which aim at minimizing the optimality gap in each round with the guarantee of user privacy and security. The details of the optimization problem are presented in the next section.
\end{itemize} 
\subsubsection{Channel noise is insufficient for protecting privacy and security with any device participation} 
In the case of $\hat{p}\leqslant p_m$, channel noise at the BS and Eve cannot guarantee qualified privacy and security for any device as an uploader if no jammer is selected. Therefore, some devices need to be selected as jammers to send jamming signals to degrade the SNR of the eavesdropper. Consequently, Policy-1 will no longer be applicable, and the only solution is Policy-2.

Then, how to choose devices that can ensure privacy and security while having a minimal negative impact on learning performance is a tradeoff problem. Therefore, we formulate an optimization problem aiming to minimize the adverse impact on the optimality gap with the consideration of privacy and security constraints in the following subsection.

\subsection{Optimization Problem of Policy-2}
In this problem we consider that a device is either selected as an uploader or as a jammer.
We introduce vector $\boldsymbol{a}=\left[ a_1,...a_N \right] $ to denote the role of devices in each round. Specifically, $a_n=1$ indicates that device $n$ is selected as an uploader, otherwise, device $n$ plays the role of a jammer. Then, the optimization problem can be formulated as follows:
\begin{align}\label{wg33}
	\mathbf{P1 }.\quad& \min_{\boldsymbol{a}} \,\,\varPsi =\frac{N\sum_{n=1}^N{\left( 1-a_n \right) p_{n,B}^{2}}+d\sigma _B}{\left( \sum_{n=1}^N{a_np_{n,B}} \right) ^2}
	\\
	\textbf{s.t.}\quad& a_n\in \left\{ 0,1 \right\} 
	 ,\forall n\in \mathcal{N}, \tag{35a}	\\ 	
	&\frac{2\kappa a_np_{n,B}}{\sqrt{\sum_{n=1}^N{\frac{\left( 1-a_n \right) p_{n,B}^{2}}{d}}+\sigma _B}}\leqslant \epsilon,\forall n\in \mathcal{N}, 
	 \tag{35b}
	\\
	&\frac{G^2}{\left( \sum_{n=1}^N{a_n} \right) ^2\underset{n}{\max}\left\{ a_np_{n,B}^{2} \right\}}\left( \sum_{n=1}^N{\frac{\left( 1-a_n \right) p_{n,E}^{2}}{d}}\,+\sigma _E\, \right) \geqslant \varUpsilon 
	. \tag{35c}
\end{align}
The objective of this problem is to minimize the impact of noise on the optimality gap as shown in (\ref{noiseimpact}). Constraint (35b) ensures privacy protection and constraint (35c) indicates the requirement of security.
Note that Problem $\mathbf{P1 }$ is a discrete nonlinear programming problem. By using the exhaustive search method (ESM), we can obtain an optimal solution to Problem $\mathbf{P1 }$. However, this method has exponential complexity. Thus, the computation cost in optimally solving Problem $\mathbf{P1 }$ is prohibitive when $N$ is large. To understand the property of the problem, we first consider a special but useful case with high-dimension learning models.

\subsection{Closed-form Solution for High-dimensional Models}
In practical scenarios, the learning model is normally with high dimensions to guarantee the learning performance. In this case, we are able to simplify the optimization problem and therefore propose closed-form optimal solutions, which could provide useful insights for practical FL systems. Assuming that $d\rightarrow \infty $, Problem $\mathbf{P1 }$ can be recast as,
\begin{align}\label{wg33}
	\mathbf{P}2.\quad& \max_{\boldsymbol{a}} \,\,\sum_{n=1}^N{a_np_{n,B}}
	\\
	\mathbf{s}.\mathbf{t}.\quad& a_n\in \left\{ 0,1 \right\} ,\forall n\in \mathcal{N} , \tag{36a}
	\\
	&\frac{2\kappa a_np_{n,B}}{\sqrt{\sigma _B}}\leqslant \epsilon ,\forall n\in \mathcal{N} ,\tag{36b}
	\\
	&\frac{G\sqrt{\sigma _E}}{\left( \sum_{n=1}^N{a_n} \right) \max \left\{ a_np_{n,B} \right\}}\geqslant \sqrt{\varUpsilon}.\tag{36c}
\end{align}
Assume that the elements in $\boldsymbol{p}_B=\left[ p_{1,B},...,p_{n,B},...,p_{N,B} \right]$ ($p_{n,B}$ is defined in Section \ref{convergenceanalysis}) are sorted in descending order. Then, we have the following result.

\begin{lemma}\label{closesolution}
	Assume that  $p_{i,B}$ is the largest one in ${\boldsymbol p}_B$ which satisfies (36b). Then, there are only $N-i+1$ closed-form solutions which may be the globally optimal solution. The $x$-th, $1\leq x \leq N-i+1$, possible solution $ {\boldsymbol{a}^x} $ is
	\begin{align}\label{wg37}
		[{\boldsymbol{a}^x}]_n=\left\{\begin{array}{l}
			1, \text { if }   i+x-1\leq n \leq  i+x+K_x-2 \\
			0, \text { otherwise }
		\end{array}\right.
	\end{align}
	where $K_x=\min \left\{ N-i-x+2,\lfloor \frac{G\sqrt{\sigma _E}}{p_{i+x-1,B}\sqrt{\Upsilon}} \rfloor \right\} $. 
\end{lemma}

\itshape {Proof:}  \upshape   Firstly, it can be found that larger number of variables that are equal to one yield a larger objective value. Therefore, we need to identify at most how many variables $a_n$ can be set to one and what indexes $n$ are they. From constraint (36b), we know that some variable $a_n$ along with large $p_{n, B}$ cannot be one since it would violate the constraint. Assuming that  $p_{i, B}$ is the largest one in ${\boldsymbol p}_B$ which satisfies (36b), it is only feasible to let $a_i,...,a_N$ equal to one.  By analyzing (36c), we can find that there are only $N-i+1$ solutions which may achieve the best performance. Specifically, the $x$-th solution corresponds to the setting that $a_1=...=a_{i+x-2}=0$ and $a_{i+x-1}=1$. In this case, from (36c), we have $\max\{a_n p_{n,B}\}=p_{i+x-1,B}$ and then the maximal number of variable $a_n$ which could be equal to one is $K_x=\min \left\{ N-i-x+2,\lfloor \frac{G\sqrt{\sigma _E}}{p_{i+x-1,B}\sqrt{\Upsilon}} \rfloor \right\}$. Then, we need to decide which $K_x$ variables are equal to one. Based on the objective function (36), clearly, the optimal allocation is to set $a_1=...=a_{i+x-2}=0$, $a_{i+x-1}=...=a_{i+x-1+K_x-1}=1$ and $a_{i+x-1+K_i}=...=a_{N}=0$.
\hfill $\blacksquare$

Based on Lemma \ref{closesolution}, we can perform the one-dimension search method to obtain the optimal solution. The optimal solution for Problem P2 is $ {\boldsymbol{a}^y} $ where
\begin{align}
	y=\arg \max _{x} \left\{   \sum_{n=1}^N     [{\boldsymbol{a}^x}]_n    p_{n, B}      \right\}.
\end{align}

 

The solution in (\ref{wg37}) proves that only a part of the variables in the middle can be set as one, which means that some devices with best and worst channel conditions cannot be selected as uploaders. This validates the trade-off of using noise between achieving privacy and security and guaranteeing learning performance. This is because the devices with the best channel conditions could result in a high risk of privacy leakage and security issue, and therefore are usually not selected as uploaders. Meanwhile, to benefit learning performance, the devices with the worst channel conditions are not usually selected as the uploader either. Inspired by the insight of the closed-form solution, we propose a heuristic algorithm based on BnB, referred to as SPA, to solve Problem $\mathbf{P1 }$, which can achieve the solution as the same as ESM with lower computational complexity.



\subsection{BnB-based SPA for Problem $\mathbf{P1 }$}

	
	
	


In the proposed algorithm, we utilize the idea of BnB to quickly cut down the branch of infeasible solutions by checking the constraints.

Assume that the elements in $\boldsymbol{p}_B=\left[ p_{1,B},...,p_{n,B},...,p_{N,B} \right]$ are sorted in the ascending order.  
It is clear that when $\sum_{n=1}^N{a_np_{n,B}}$ is small, the value of objective function is large. By contrast, it can be observed from constraints (35b) and (35c) that the fewer the number of $a_n=1,n\in \mathcal{N}$ and the smaller the $p_{n, B}$ are, the easier the constraints can be satisfied. More specifically, if $\boldsymbol{a}=\left[ 1,0,...,0,...,0 \right]$ cannot satisfy constraints (35b) and (35c), any other solutions $\boldsymbol{a}=\left[ 1,a_2,...,a_n,...,a_N \right] $ cannot meet constraints (35b) and (35c) either. In this context, all the solutions with $a_1 = 1$ are infeasible and should be discarded. Following this idea, we can delete half of the solution space of the subproblem in each branch-and-bound round, and therefore, we can keep narrowing the search space effectively. Given the property of the objective function, we try to get the solution with more variables equal to 1 while satisfying the constraints. Besides, to introduce more diversity to the solutions, we will branch and bound starting from the different indexes of the nodes, i.e., from $a_n, \forall n$. Specifically, one round of the detailed branch-and-bound process from $a_n$ is described as follows:

\begin{itemize}
	\item Branching: Select the current node $a_n$ that has not been not branched yet. We branch it into two nodes: one is to set it as the uploader, and the other is to set it as the jammer.
	
	\item Bounding: Check if $\boldsymbol{a}=\left[ 0,...,0,a_n=1,0,...,0 \right] $ meets the constraints (35b) and (35c).
	
		\item Pruning: If $\boldsymbol{a}=\left[ 0,...,0,a_n=1,0,...,0 \right] $ satisfies constraints (35b) and (35c), the node is selected as an uploader since this selection scheme would definitely lead to a better objective value than selecting this node as a jammer.
		On the other word, the branch with $a_n=0$ is cut off. %
Otherwise, this node is selected as jammer and the branch with $a_n=1$ is cut off.
\end{itemize} 




By defining $\boldsymbol{q}_B=\left[ p_{1,B}^{2},...,p_{n,B}^{2},...,p_{N,B}^{2} \right]$, $\boldsymbol{q}_E=\left[ p_{1,E}^{2},...,p_{n,E}^{2},...,p_{N,E}^{2} \right]$ and iteratively conducting the branch-and-bound process, the overall algorithm for solving Problem $\mathbf{P1}$ is formally presented in Algorithm 1. 
\begin{algorithm}[ht]  
	\caption{BnB-based SPA Algorithm for Solving Problem $\mathbf{P1 }$}  
	\label{Iteration}  
	\begin{algorithmic}[1]  
		\Require 
		Given $N$, $d$, $G$, $\sigma_B$, $\sigma_E$, $\boldsymbol{p}_B$, $\boldsymbol{q}_B$, $\boldsymbol{q}_E$, $\epsilon$ and $\varUpsilon$. Initialize $\varPsi ^*=+\infty $.
		\Ensure  
		$\boldsymbol{a}^*$
		\For{$iter\in \left[ 1,N \right] $}
		\State Let $\boldsymbol{a}^{\left( iter \right)}=\left[ 0,...,0,...,0 \right]$.
			\For {$index\in \left[ iter,N \right]$}
			\State Let $a_{index}^{\left( iter \right)} = 1$.
			 \If {$\boldsymbol{a}^{\left( iter \right)}$ is infesiable for (35b) and (35c)}
			 \State 
			 Set $a_{index}^{\left( iter \right)} = 0$.
			 \EndIf
			\EndFor
	\State Compute the objective value $\varPsi \left( \boldsymbol{a}^{\left( iter \right)} \right)$.
	\If {$\varPsi \left( \boldsymbol{a}^{\left( iter \right)} \right) \leqslant \varPsi ^*$}
	\State $\varPsi ^*=\varPsi \left( \boldsymbol{a}^{\left( iter \right)} \right)$ and $\boldsymbol{a}^*=\boldsymbol{a}^{\left( iter \right)}$.
	\EndIf	
	\EndFor
		\end{algorithmic}  
\end{algorithm}  



\section{Simulation Results}\label{simulation}
In this section, we evaluate the performance of the proposed SPA algorithm, the scheduling policies and the CWPP mechanism. In the case that channel noise is sufficient for the private and secure participation of full devices, we just schedule all the devices to participate in training. Therefore, we do not consider the simulation of this case. We first introduce the general simulation setting in Subsection \ref{simulationsetting}. Then, we evaluate the performance of the proposed BnB-based SPA algorithm for Policy-2 in Subsection \ref{performanceofBnb} and the time-complexity in Subsection \ref{timecomplexityofBnb}. The comparison of the performance between Policy-1 and Policy-2 is plotted in Subsection \ref{Policy1andPolicy2}. We finally evaluate the proposed post-processing mechanism in Subsection \ref{performanceofCWPP}.

\subsection{Simulation Setting}\label{simulationsetting}
We assume that the wireless channels from edge devices to the BS and Eve follow Rayleigh distribution in different communication rounds. 
We evaluate our proposed scheme by training a convolutional neural network (CNN) on the popular MNIST \cite{lecun2010mnist} dataset used for handwritten digit classification. The MNIST dataset consists of 60,000 images for training and 10,000 testing images of the 10 digits.
We have the general assumption that there is an equal number of training data samples for each device and no overlap between the local training data sets \cite{zhu2019broadband} \cite{wang2019adaptive}.
We assume that local datasets are IID, where the initial training dataset is randomly divided into $N$ batches and each device is assigned to one batch. In particular, CNN consists of two 5×5 convolution layers with the rectified linear unit (ReLU) activation. The two convolution layers have 10 and 20 channels respectively, and each layer has 2×2 max pooling, a fully-connected layer with 50 units and ReLU activation, and a log-softmax output layer, in which case $d= 21840$. The learning rate is set as $\eta=0.1$.

\subsection{Evaluation of SPA for Ploicy-2}\label{performanceofBnb}
In this section, we evaluate the performance of the proposed BnB-based SPA algorithm in solving the optimization problem in Policy-2 by comparing it with the genetic algorithm (GA), ESM, and random solution. In GA settings, we utilize the python tools named geatpy. The transmit power budgets at each device are assumed to be the same and are set to $P_k=5W$. Both the powers of the additive Gaussian noise at the BS and Eve are set to $\sigma _B=\sigma _E=1$. The privacy and security coefficients are $\epsilon =12$ and $\varUpsilon =1.5$, respectively.

\begin{figure*}[ht]
	\centering
	\subfigure[$N=25$]	
	{  \begin{minipage}[t]{0.48\linewidth}
			\centering
			\includegraphics[scale=0.8]{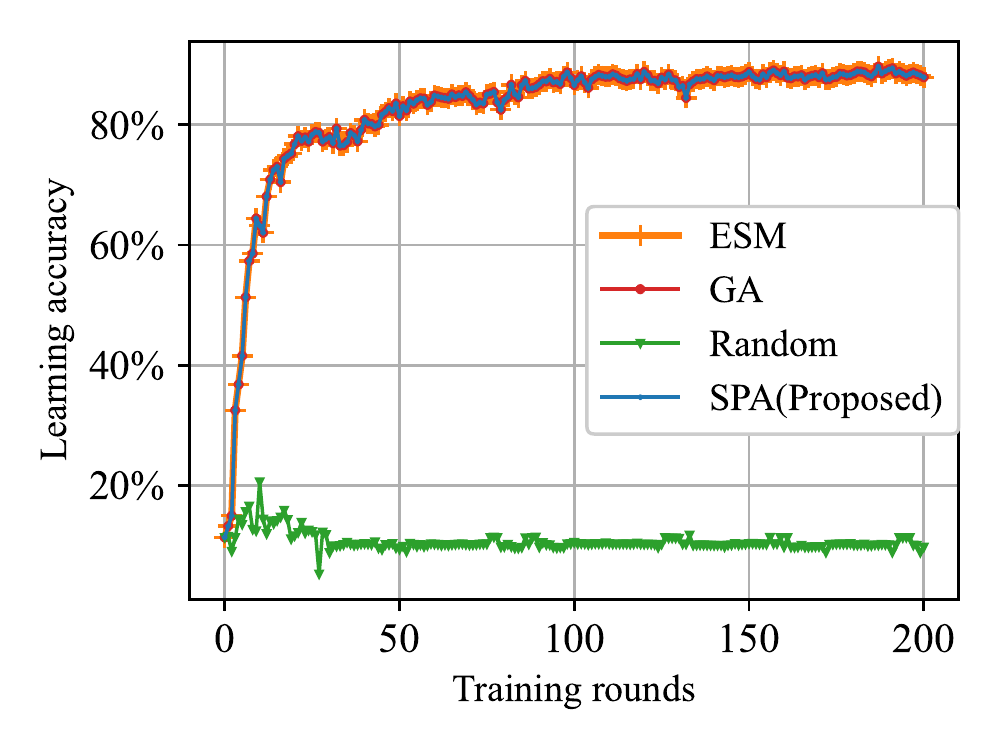}
		\end{minipage}	
	}
	\subfigure[$N=100$]	
	{  \begin{minipage}[t]{0.48\linewidth}
			\centering
			\includegraphics[scale=0.8]{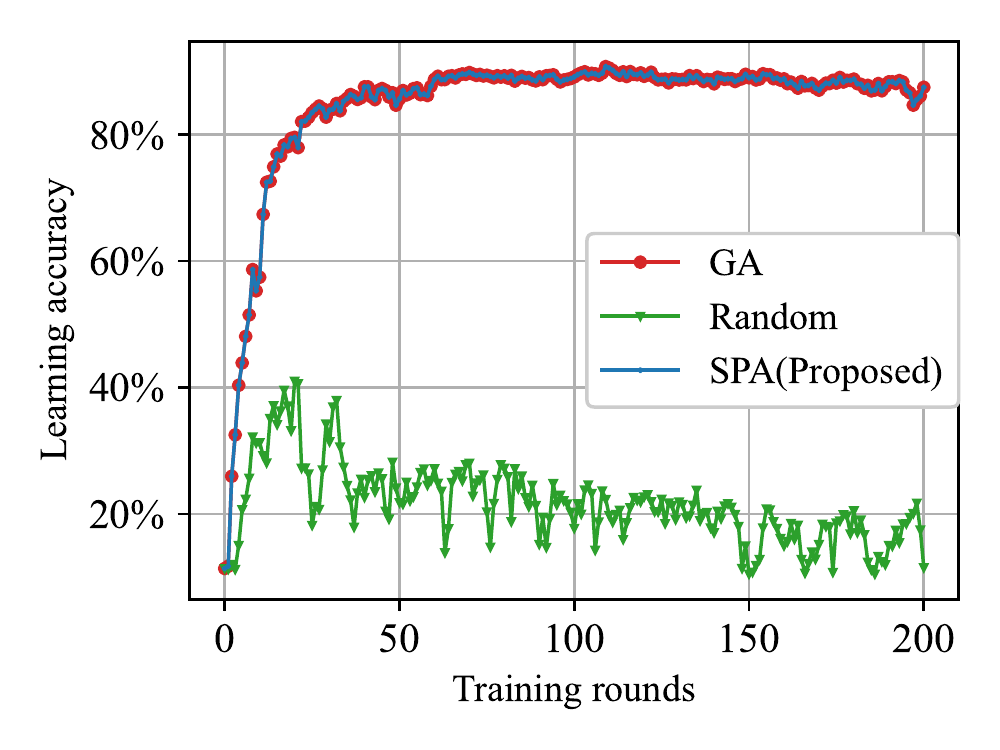}
		\end{minipage}	
	}
	\caption{The learning performance with different solving methods} \label{simulation1}
\end{figure*}
In Fig. \ref{simulation1}, we illustrate the learning accuracy under different device decisions which are obtained by applying SPA, ESM, GA and random scheduling. Particularly, there is no doubt that ESM can obtain the optimal solution, however, we only employ ESM in case of $K=25$ due to the extra high time complexity when $N$ is large. In Fig. \ref{simulation1} (a), the result validates that SPA and GA can achieve the same learning performance as ESM, which means that the SPA and GA could obtain the same optimal solution as ESM. When $N$ becomes larger as shown in Fig. \ref{simulation1} (b), the proposed SPA can still achieve the same performance as GA, which demonstrates that the proposed SPA is effective to achieve the optimal solution even with the large $N$.

\begin{figure}[ht]
	\centering
	\includegraphics[scale=0.7]{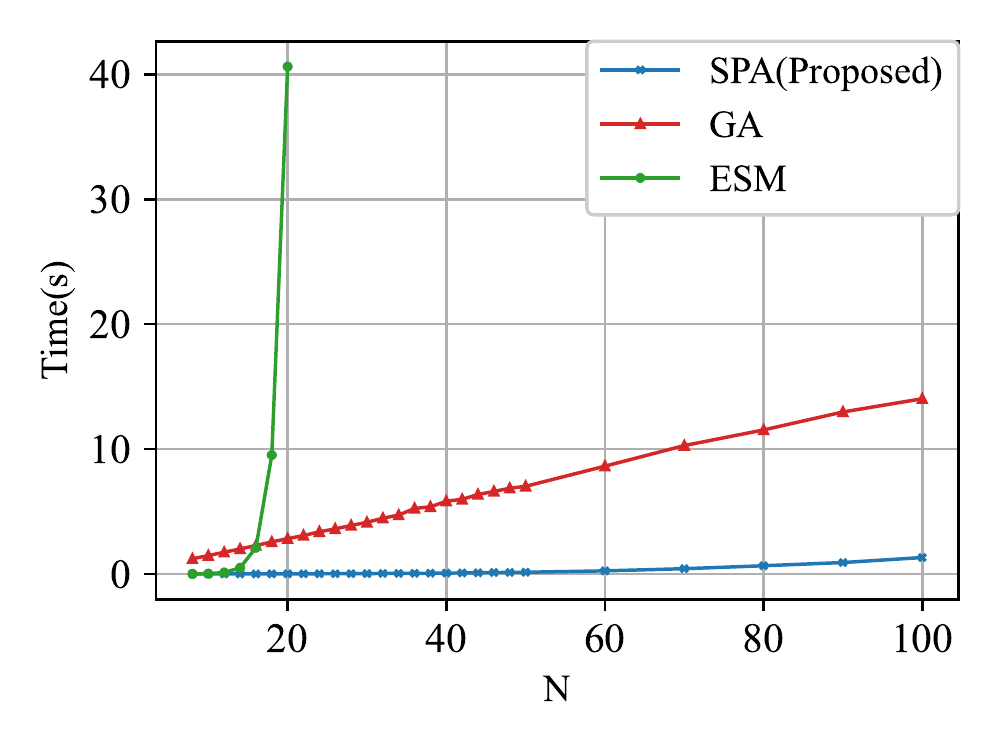}
	\caption{Time comparasion-eps-converted-to} \label{simulation2}
\end{figure}
\subsection{Evaluation of Time-complexity of SPA}\label{timecomplexityofBnb}
In Fig. \ref{simulation2}, we plot the execution time of the proposed SPA algorithm, GA, and ESM where the security and privacy coefficients are set as $\varUpsilon =1.5$ and $\epsilon =12$, respectively. The results reveal that in the case of relatively small $N$, i.e., 8,10,12, the execution time of ESM is similar to the SPA and less than that of GA. However, as $N$ increases from $N=14$, the time consumed by ESM increases sharply, therefore, the time consumption of ESM is omitted when $N$ is large than 20. By contrast, the proposed SPA always consumes the least amount of time and the growth rate is slow as $N$ increases. Therefore, the proposed SPA can achieve the same performance as ESM while maintaining a very low complexity, indicating that it is promising to be applied in large-scale FL systems.
\subsection{Evaluation of Ploicy-1 and Ploicy-2}\label{Policy1andPolicy2}

\begin{figure*}[ht]
	\centering
	\subfigure[$N=25$]	
	{  \begin{minipage}[t]{0.31\linewidth}
			\centering
			\includegraphics[scale=0.6]{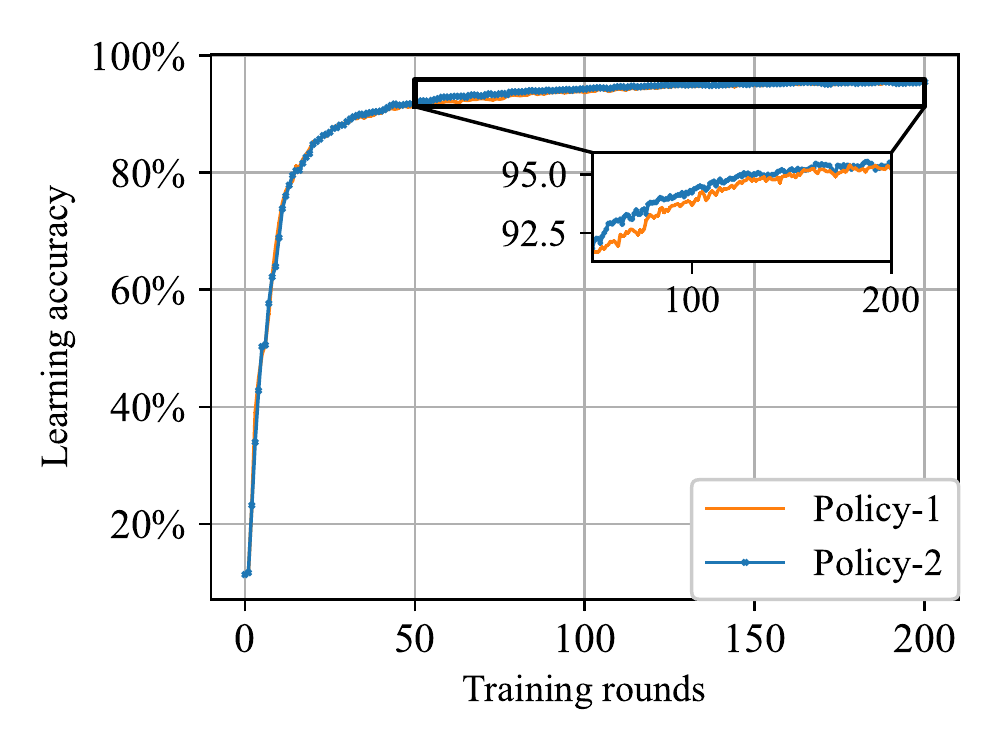}
		\end{minipage}	
	}
	\subfigure[$N=50$]	
	{  \begin{minipage}[t]{0.31\linewidth}
			\centering
			\includegraphics[scale=0.6]{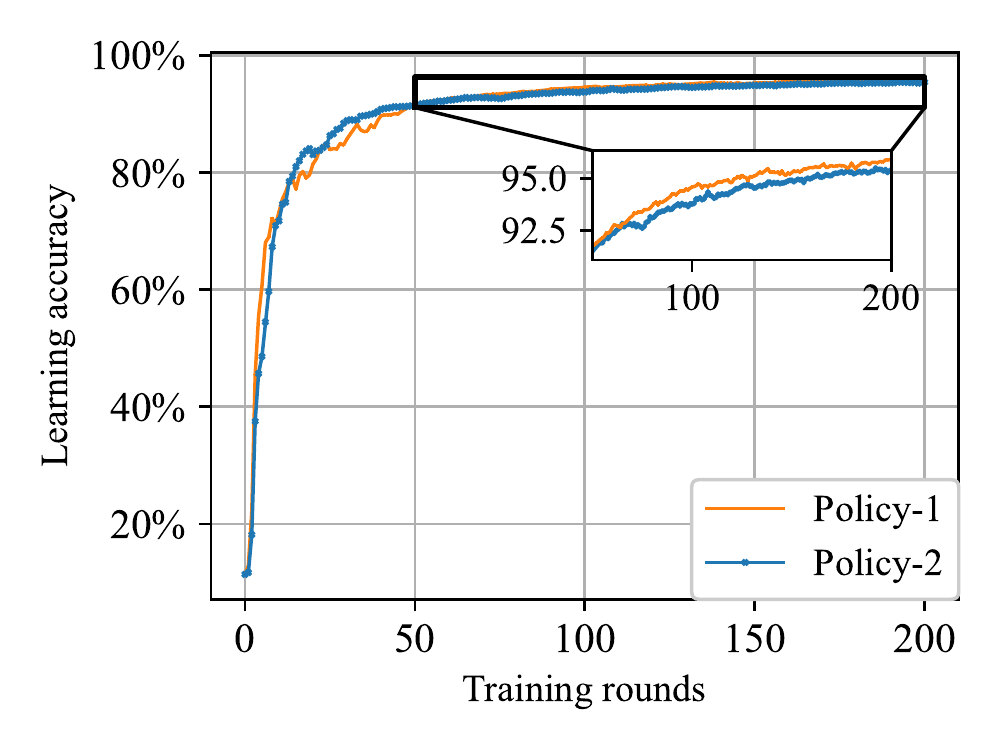}
		\end{minipage}	
	}
	\subfigure[$N=100$]	
	{  \begin{minipage}[t]{0.31\linewidth}
			\centering
			\includegraphics[scale=0.6]{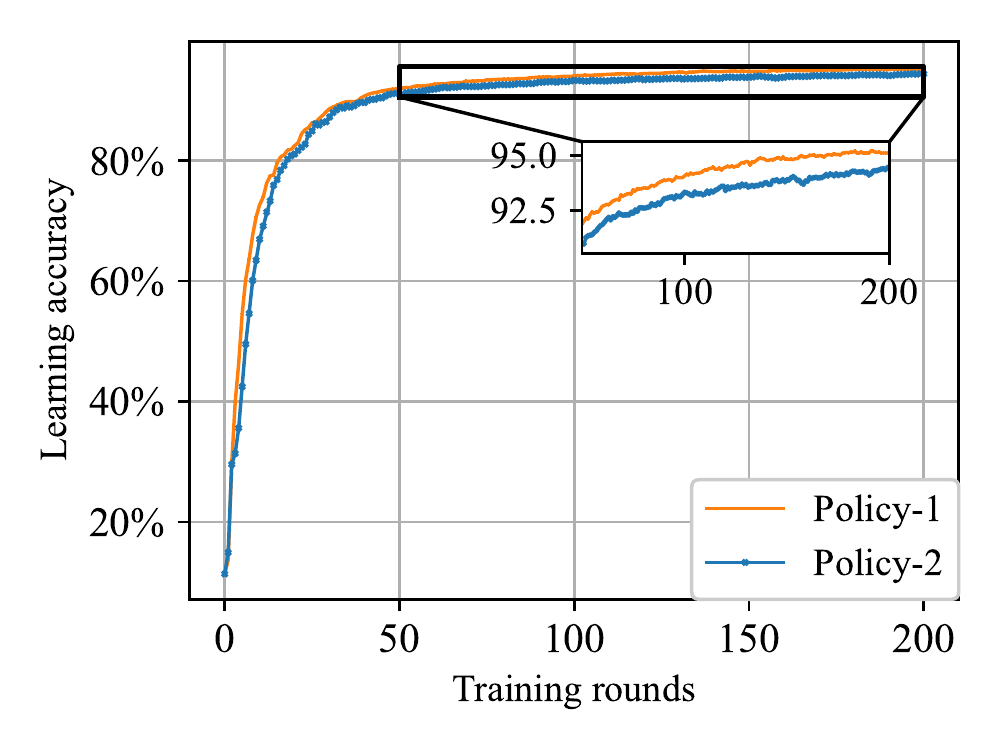}
		\end{minipage}	
	}

	\caption{Comparison of learning performance between Policy-1 and Ploicy-2} \label{simulation3}
\end{figure*}
We evaluate the performance of Policy-1 and Policy-2 in Fig. \ref{simulation3} and study the impact of the number of devices and power budget on the performance where the security and privacy coefficients are set to $\varUpsilon =0.5$ and $\epsilon =20$, respectively. The power budgets of each devices are set to $P=5W$.

In Fig. \ref{simulation3}, we can observe that as the number of devices increases, Policy-1 gradually performs better than Policy-2.
It is well known that more participants and less noise distortion can make the model more accurate.
The drawback of Policy-1 is that the number of selected devices to participate in training is less than Policy-2. Due to the introduction of artificial noise, Policy-2 results in more noise distortion in the training process, which makes the model less accurate. As the total number of devices becomes larger, more devices are qualified to be selected as participants in Policy-1 where the noise only contains channel noise. By contrast, although there are more devices selected to participate in training in Policy-2, the power of noise is also bigger than that in Policy-1. Therefore, Policy-1 performs better than Policy-2 in the case when the number of devices is bigger.

\subsection{Evaluation of CWPP Mechanism in case of Sufficient Channel Noise} \label{performanceofCWPP}
\begin{figure*}[ht]
	\centering
	\subfigure[$K=25$]	
	{  \begin{minipage}[t]{0.31\linewidth}
			\centering
			\includegraphics[scale=0.6]{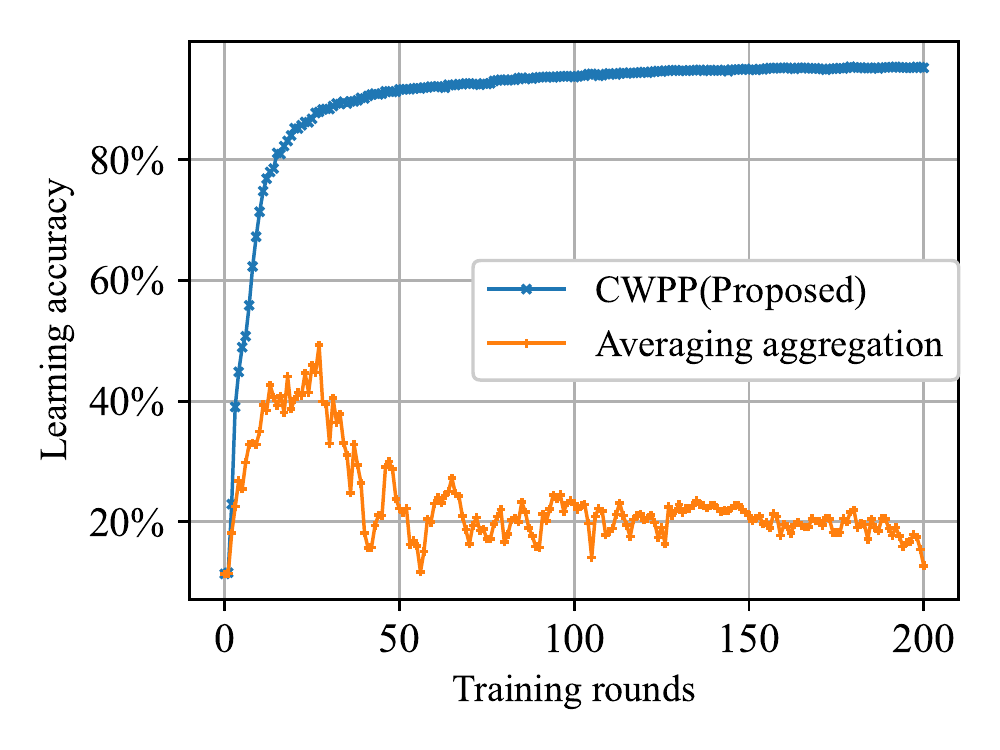}
		\end{minipage}	
	}
	\subfigure[$K=50$]	
	{  \begin{minipage}[t]{0.31\linewidth}
			\centering
			\includegraphics[scale=0.6]{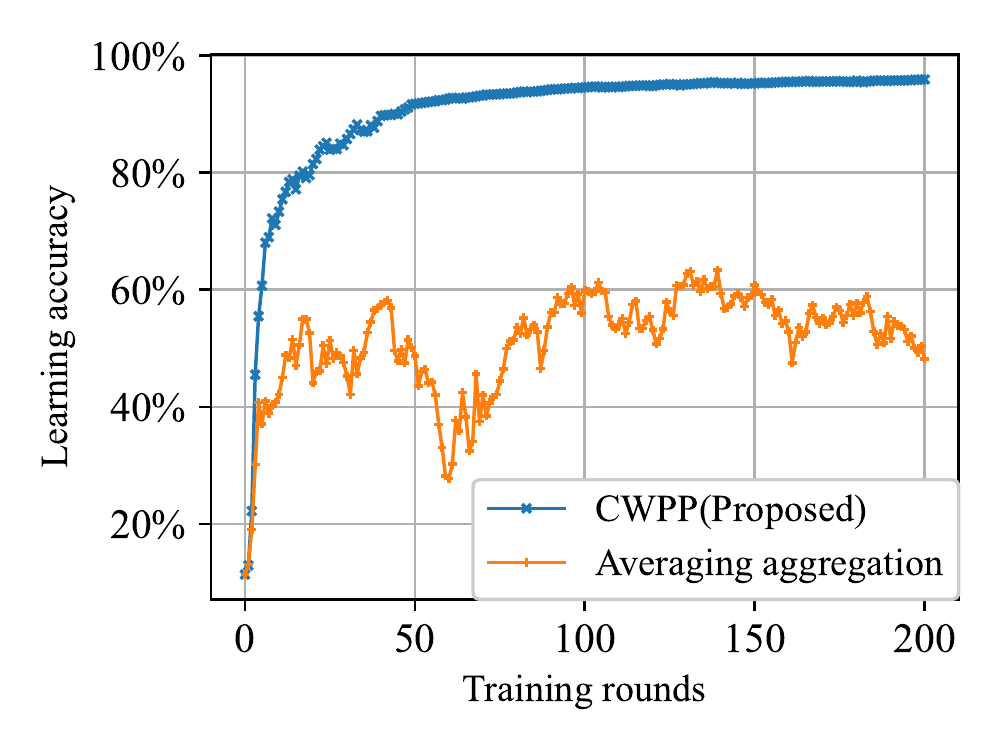}
		\end{minipage}	
	}
	\subfigure[$K=100$]	
	{  \begin{minipage}[t]{0.31\linewidth}
			\centering
			\includegraphics[scale=0.6]{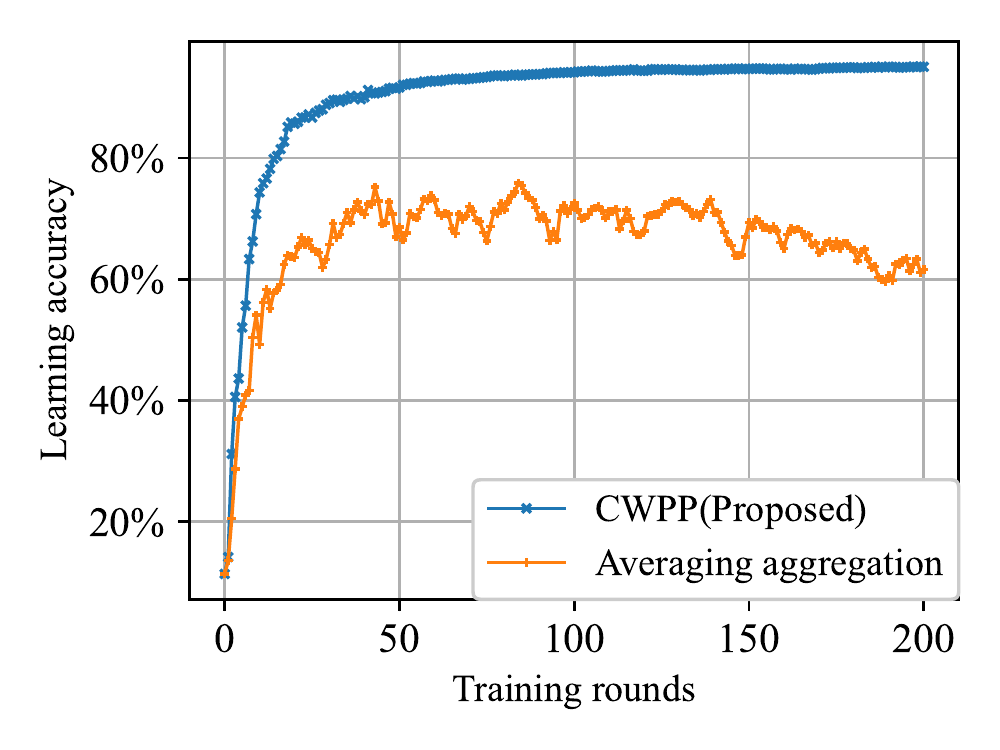}
		\end{minipage}	
	}
	\caption{Comparison of learning performance between CWPP and averaging aggregation} \label{simulation4}
\end{figure*}
We evaluate the performance of the proposed CWPP mechanism by comparing it with the averaging OTA post-processing mechanism \cite{seif2020wireless}.
In the averaging post-processing mechanism, which has been referred to as aligned OTA-FL in our previous work \cite{yan2022private}, the gradients need to be aligned by an alignment coefficient, i.e., $c=\underset{n\in \mathcal{K}}{\min}\left\{ h_{n,B}P_n \right\}$. However, in reality, most edge devices are low-powered, which will result in a quite small alignment coefficient. Therefore, the SNR of the system will be degraded to a quite low level and the aggregated gradients will be less accurate. A more detailed analysis can be found in \cite{yan2022private}.

For clarity, we compare the two post-processing mechanisms with Policy-1 device selection algorithm where the aggregated noise only contains channel noise and is independent of the channel quality of the selected uploaders. The security and privacy coefficients are set to $\varUpsilon =0.5$ and $\epsilon =20$, respectively. 

The results in Fig. \ref{simulation4} validate that the proposed CWPP mechanism is superior to the averaging post-processing. First of all, the CWPP scheme avoids gradient alignment, and therefore the SNR of other devices will not be limited by the devices with poor channel conditions. Additionally, from Equation (\ref{channel-weighted-2}), we can learn that the BS assigns a bigger weight to the gradient from the device with better channel conditions, and thereby, mitigating the negative impact of noise on the learning process.

\section{Conclusion}\label{conclusion}
To enhance the privacy of user data and the security of the FL system, we have proposed an SP-OTA-FL framework in this work. In the proposed FL system, noise is used to protect both user privacy and system security. Specifically, three cases have been considered. In particular, we have quanitified the privacy leakage of user data using DP, and measured the security of the analog gradient using MSE-security. To reduce the impact of noise on the aggregated gradient, we have proposed a CWPP mechanism, which assigns less weight to badly distorted gradients. 
Further, we have conducted the privacy, security and convergence analysis and theoretically characterized the impacts of noise on privacy and security protection as well as the optimality gap. To obtain an appropriate device selection decision, we have analyzed the proposed framework and proposed two policies for device selection when the channel noise is insufficient. In particular, we have formulated an integer fractional optimization problem, which can be solved with low complexity via a BnB-based SPA algorithm. The effectiveness of the proposed CWPP and the device selection policies has been validated through simulation.

\appendices
\section{Proof of Lemma \ref{lemma1}}\label{proofofsmoothness}
The proof of smoothness is given as follows:
\begin{equation}\label{wg15}
	\begin{aligned}
		&L\left( \boldsymbol{\iota }' \right) -L\left( \boldsymbol{\iota } \right) \overset{\left( a \right)}{=}\frac{1}{N}\sum_{n=1}^N{L_n\left( \boldsymbol{\iota }' \right)}-\frac{1}{N}\sum_{n=1}^N{L_n\left( \boldsymbol{\iota } \right)}
		\\
		\overset{\left( b \right)}{\leqslant}&\frac{1}{N}\sum_{n=1}^N{\left[ \left( \boldsymbol{\iota }'-\boldsymbol{\iota } \right) ^{\mathrm{T}}\nabla L_n\left( \boldsymbol{\iota } \right) +\frac{\theta}{2}\left\| \boldsymbol{\iota }'-\boldsymbol{\iota } \right\| _{2}^{2} \right]}
		\\
		=&\left( \boldsymbol{\iota }'-\boldsymbol{\iota } \right) ^{\mathrm{T}}\left( \frac{1}{N}\sum_{n=1}^N{\nabla L_n\left( \boldsymbol{\iota } \right)} \right) +\frac{\theta}{2}\left\| \boldsymbol{\iota }'-\boldsymbol{\iota } \right\| _{2}^{2}
		\\
		=&\left( \boldsymbol{\iota }'-\boldsymbol{\iota } \right) ^{\mathrm{T}}\nabla L\left( \boldsymbol{\iota } \right) +\frac{\theta}{2}\left\| \boldsymbol{\iota }'-\boldsymbol{\iota } \right\| _{2}^{2},
	\end{aligned}
\end{equation}
where (a) is from the definition of $L\left( \cdot \right)$ in (\ref{globallossfunction}) and (b) is from Assumption \ref{assumption1}. The proof of convexity has a similar process and is therefore omitted here. 

\section{Proof of Lemma \ref{lemma2}}\label{proofofprivacyanalysis}
Here we use index $k$ instead of $n$ to avoid confusion between the specific index of device $n$ and the notation $n$ in the summation. 

Assume that $\mathcal{B} _k$ and $\mathcal{B} _{k}^{'}$ are two adjacent datasets differing in one sample. $\left( \boldsymbol{y}^t \right) '$ is the received signal at the BS, which only differs in one gradient with $\boldsymbol{y}^t$. The gradient $\left( \boldsymbol{g}_{k}^{t} \right) '$ from uploader $k$ in $\left( \boldsymbol{y}^t \right) '$ is obtained based on $\mathcal{B} _{k}^{'}$.
Based on the definition of sensitivity and Assumption \ref{gradientassumption}, one has
\begin{equation}\label{wg15}
	\begin{aligned}
		\varDelta S_{k}^{t}\triangleq& \underset{\mathcal{B} _k,\mathcal{B} _{k}^{'}}{\max}\left\| \boldsymbol{y}^t-\left( \boldsymbol{y}^t \right) ' \right\| _2
		\\
		=&\underset{\mathcal{B} _n,\mathcal{B} _{k}^{'}}{\max}\left\| \frac{h_{k,B}^{t}\sqrt{P_k}}{G}\left( \boldsymbol{g}_{k}^{t}-\left( \boldsymbol{g}_{k}^{t} \right) ' \right) \right\| _2
		\\
		=&\frac{h_{k,B}^{t}\sqrt{P_k}}{G}\left\| \boldsymbol{g}_{k}^{t}-\left( \boldsymbol{g}_{k}^{t} \right) ' \right\| _2\overset{\left( a \right)}{\leqslant}2h_{k,B}^{t}\sqrt{P_k},
	\end{aligned}
\end{equation}
where (a) is from Triangular Inequality and Assumption \ref{gradientassumption}.
In accordance with the Gaussian mechanism of DP and the above result, one completes the proof of Lemma \ref{lemma2}.

\section{Proof of Lemma \ref{securitylemma}}\label{proofofsecuritylemma}

Firstly, since the elements in $\boldsymbol{g}_{n}^{t}$ are uniformly distributed in $\left[ a,b \right] $, the $\boldsymbol{g}_{ave}^{t}$ follows the same distribution in $\left[ a,b \right]$. 
For analysis, we define $\tilde{\mathcal{E}}^t:\left( \boldsymbol{g}_{n}^{t} \right) _{n\in \mathcal{K} ^t}\rightarrow \boldsymbol{\tilde{z}}^t\in \tilde{\mathcal{Z}}$ where 
\begin{equation}\label{securitylemmawg2}
	\begin{aligned}
		\boldsymbol{\tilde{z}}^t=\sum_{n\in \mathcal{K} ^t}{\frac{\varLambda ^t}{G}\boldsymbol{g}_{n}^{t}}+\boldsymbol{r}_{E,Tot}^{t}.
	\end{aligned}
\end{equation}
Assume that the variance of $\boldsymbol{\tilde{z}}^t$ is $\sigma$.
Following Lemma 3 and Lemma 4 in \cite{frey2020towards}, the minimum MSE estimator $e\left( \boldsymbol{\tilde{z}}^t \right)$ for estimating $\boldsymbol{g}_{ave}^{t}$ from the observations $\boldsymbol{\tilde{z}}^t$ satisfies:
\begin{equation}\label{wg28}
	\begin{aligned}
		\mathbb{E} \left[ \left( \boldsymbol{g}_{ave}^{t}-e\left( \boldsymbol{\tilde{z}}^t \right) \right) ^2 \right] =\sigma \varXi \left( \frac{b-a}{\sqrt{\sigma}} \right).
	\end{aligned}
\end{equation}
The lowest-variance unbiased estimator is:
\begin{equation}\label{securitylemmawg2}
	\begin{aligned}
		e\left( \boldsymbol{\tilde{z}}^t \right) =\boldsymbol{g}_{ave}^{t}+\frac{G}{\left| \mathcal{K} \right|^t\varLambda ^t}\boldsymbol{r}_{E,Tot}^{t},
	\end{aligned}
\end{equation}
with the variance given by
\begin{equation}\label{securitylemmawg2}
	\begin{aligned}
		\gamma _{E}^{t}=\frac{G^2}{\left( \left| \mathcal{K} \right|^t\varLambda ^t \right) ^2}\left( \sum_{n\in \mathcal{J} ^t}{\frac{\left( h_{n,E}^{t} \right) ^2P_n}{d}}+\sigma _E\,\, \right),
	\end{aligned}
\end{equation}	
where $\varLambda ^t=\underset{n\in \mathcal{K} ^t}{\max}\left\{ h_{n,B}^{t}\sqrt{P_n} \right\} $. It thus follows from (\ref{wg28}) and Definition \ref{difinition3} that $\tilde{\mathcal{E}}^t$ guarantees $\left( \tilde{\mathcal{E}}^t,\gamma _{E}^{t}\varXi \left( \frac{b-a}{\sqrt{\gamma _{E}^{t}}} \right) \right)$.

On the other hand, one has
\begin{equation}\label{securitylemmawg2}
	\begin{aligned}
		\mathbb{E} \left[ \left\| e\left( \boldsymbol{\tilde{z}}^t \right) -\boldsymbol{g}_{ave}^{t} \right\| ^2 \right] =\left( \frac{G}{\left| \mathcal{K} \right|^t\varLambda ^t} \right) ^2\mathbb{E} \left[ \left\| \boldsymbol{r}_{E,Tot}^{t} \right\| ^2 \right].
	\end{aligned}
\end{equation}	
Similarly, we also have
\begin{equation}\label{securitylemmawg2}
	\begin{aligned}
		&\mathbb{E} \left[ \left\| e\left( \boldsymbol{z}^t \right) -\boldsymbol{g}_{ave}^{t} \right\| ^2 \right] 
		\\
		=&\mathbb{E} \left[ \left\| \sum_{n\in \mathcal{K} ^t}{\left( \frac{h_{n,E}^{t}\sqrt{\lambda _{n}^{t}}}{\,\,\varLambda ^t}-1 \right) \boldsymbol{g}_{n}^{t}}+\frac{G}{\varLambda ^t}\boldsymbol{r}_{E,Tot}^{t} \right\| ^2 \right] 
		\\
		\overset{\left( a \right)}{=}&\frac{1}{\left| \mathcal{K} ^t \right|^2}\mathbb{E} \left[ \left\| \sum_{n\in \mathcal{K} ^t}{\left( \frac{h_{n,E}^{t}\sqrt{\lambda _{n}^{t}}}{\,\,\varLambda ^t}-1 \right) \boldsymbol{g}_{n}^{t}} \right\| ^2 \right] 
		+\left( \frac{G}{\left| \mathcal{K} \right|^t\varLambda ^t} \right) ^2\mathbb{E} \left[ \left\| \boldsymbol{r}_{E,Tot}^{t} \right\| ^2 \right] ,
	\end{aligned}
\end{equation}
where (a) comes from $\mathbb{E} \left[ \boldsymbol{r}_{E,Tot}^{t} \right] =0$. Obviously, $\mathbb{E} \left[ \left\| e\left( \boldsymbol{\tilde{z}}^t \right) -\boldsymbol{g}_{ave}^{t} \right\| ^2 \right]$ is smaller than $\mathbb{E} \left[ \left\| e\left( \boldsymbol{z}^t \right) -\boldsymbol{g}_{ave}^{t} \right\| ^2 \right]$, therefore, $e\left( \boldsymbol{\tilde{z}}^t \right)$ is a closer estimate of $\boldsymbol{g}_{ave}^{t}$. Then, $ e\left( \boldsymbol{z}^t \right)$ has a larger variance and can achieve at least  $\left( \tilde{\mathcal{E}}^t,\gamma _{E}^{t}\varXi \left( \frac{b-a}{\sqrt{\gamma _{E}^{t}}} \right) \right) $-MSE-security.

Alternatively, from the communication point of view, one can also get that $\boldsymbol{\tilde{z}}^t$ could have a better recovery of gradient than $\boldsymbol{z}^t$ because of a higher SNR as $\varLambda ^t=\underset{n\in \mathcal{K} ^t}{\max}\left\{ h_{n,B}^{t}\sqrt{P_n} \right\} $. Therefore, if $\boldsymbol{\tilde{z}}^t$ can guarantee at least $\left( \tilde{\mathcal{E}}^t,\gamma _{E}^{t}\varXi \left( \frac{b-a}{\sqrt{\gamma _{E}^{t}}} \right) \right) $-MSE-security, then so can $\boldsymbol{z}^t$ . Then, we complete the proof of Lemma \ref{securitylemma}.

\section{Proof of Lemma \ref{lemma4}}\label{proofoflemma4} 
Based on the definitions in (\ref{noiselessSGD}) and Assumption 1, one has
\begin{equation}\label{wg52}
	\begin{aligned}
		\mathbb{E} \left[ \boldsymbol{\hat{g}}^t-\boldsymbol{\bar{g}}^t \right] =\sum_{n\in \mathcal{K} ^t}{\frac{p_{n,B}^{t}}{\sum_{n\in \mathcal{K} ^t}{p_{n,B}^{t}}}\mathbb{E} \left[ \boldsymbol{g}_{n}^{t}-\nabla L_n\left( \boldsymbol{m}_{n}^{t} \right) \right]}=0.
	\end{aligned}
\end{equation}
Then, we also have,
\begin{equation}\label{wg52}
	\begin{aligned}
		&\mathbb{E} \left[ \left\| \boldsymbol{\hat{g}}^t-\boldsymbol{\bar{g}}^t \right\| _{2}^{2} \right] 
		\\
		=&\mathbb{E} \left[ \left\| \sum_{n\in \mathcal{K} ^t}{\frac{p_{n,B}^{t}}{\sum_{n\in \mathcal{K} ^t}{p_{n,B}^{t}}}\left( \boldsymbol{g}_{n}^{t}-\nabla L_n\left( \boldsymbol{m}_{n}^{t} \right) \right)} \right\| _{2}^{2} \right] 
		\\
		\overset{\left( a \right)}{\leqslant}&\sum_{n\in \mathcal{K} ^t}{\frac{p_{n,B}^{t}}{\sum_{n\in \mathcal{K} ^t}{p_{n,B}^{t}}}\mathbb{E} \left[ \left\| \boldsymbol{g}_{n}^{t}-\nabla L_n\left( \boldsymbol{m}_{n}^{t} \right) \right\| _{2}^{2} \right]}
		\\
		\overset{\left( b \right)}{\leqslant}&\vartheta ^2,
	\end{aligned}
\end{equation}
where (a) is obtained by using Jensen's inequality and (b) is from (\ref{wg18}). We therefore complete the proof of Lemma \ref{lemma4}.

\section{Proof of Theorem 1}\label{proofofoptimalitygap} 
Accroding to the update process of the global model shown in (\ref{globalmodeupdate}), the gap between the global model parameter $\boldsymbol{m}^{t+1}$ and the optimal global model parameter $\boldsymbol{m}^*$ can be expressed as,

\begin{equation}\label{wg45}
	\begin{aligned}
		&\mathbb{E} \left[ \left\| \boldsymbol{m}^{t+1}-\boldsymbol{m}^* \right\| _{2}^{2} \right] 
		\\
		=&\mathbb{E} \left[ \left\| \boldsymbol{m}^t-\tau ^t\boldsymbol{\hat{g}}^t-\tau ^t\frac{G}{\sum_{n\in \mathcal{K} ^t}{p_{n,B}^{t}}}\boldsymbol{r}_{B,Tot}^{t}-\boldsymbol{m}^* \right\| _{2}^{2} \right] 
		\\
		=&\mathbb{E} \left[ \left\| \boldsymbol{m}^t-\tau ^t\boldsymbol{\hat{g}}^t+\tau ^t\boldsymbol{\bar{g}}^t-\tau ^t\boldsymbol{\bar{g}}^t \right. \right. \left. \left. -\tau ^t\frac{G}{\sum_{n\in \mathcal{K} ^t}{p_{n,B}^{t}}}\boldsymbol{r}_{B,Tot}^{t}-\boldsymbol{m}^* \right\| _{2}^{2} \right] 
		\\
		\overset{\left( a \right)}{=}&\mathbb{E} \left[ \left\| \boldsymbol{m}^t-\tau ^t\boldsymbol{\bar{g}}^t-\tau ^t\frac{G}{\sum_{n\in \mathcal{K} ^t}{p_{n,B}^{t}}}\boldsymbol{r}_{B,Tot}^{t}-\boldsymbol{m}^* \right\| _{2}^{2} \right] +\left( \tau ^t \right) ^2\mathbb{E} \left[ \left\| \boldsymbol{\hat{g}}^t-\boldsymbol{\bar{g}}^t \right\| _{2}^{2} \right] 
		\\
		\overset{\left( b \right)}{=}&\mathbb{E} \left[ \left\| \boldsymbol{m}^t-\tau ^t\boldsymbol{\bar{g}}^t-\boldsymbol{m}^* \right\| _{2}^{2} \right] +\left( \frac{\tau ^tG}{\sum_{n\in \mathcal{K} ^t}{p_{n,B}^{t}}} \right) ^2\mathbb{E} \left[ \left\| \boldsymbol{r}_{B,Tot}^{t} \right\| _{2}^{2} \right] +\left( \tau ^t \right) ^2\mathbb{E} \left[ \left\| \boldsymbol{\hat{g}}^t-\boldsymbol{\bar{g}}^t \right\| _{2}^{2} \right] 
		\\
		\overset{\left( c \right)}{=}&\mathbb{E} \left[ \left\| \boldsymbol{m}^t-\boldsymbol{m}^* \right\| _{2}^{2} \right] +\underset{A}{\underbrace{\left( \tau ^t \right) ^2\mathbb{E} \left[ \left\| \boldsymbol{\bar{g}}^t \right\| _{2}^{2} \right] }}\underset{B}{\underbrace{-2\tau ^t\left< \boldsymbol{m}^t-\boldsymbol{m}^*,\boldsymbol{\bar{g}}^t \right> }}+\left( \tau ^t \right) ^2\vartheta ^2
		\\&+\left( \frac{\tau ^tG}{\sum_{n\in \mathcal{K} ^t}{p_{n,B}^{t}}} \right) ^2\underset{C}{\underbrace{\mathbb{E} \left[ \left\| \boldsymbol{r}_{B,Tot}^{t} \right\| _{2}^{2} \right] }},
	\end{aligned}
\end{equation}
where (a) and (c) are obtained by applying Lemma \ref{lemma4}. Step (b) is from the fact that $\mathbb{E} \left[ \boldsymbol{r}_{B,Tot}^{t} \right] =0$.

Then, we obtain the upper bounds for each term in (\ref{wg45}), separately. Firstly, we have the upper bound of term $A$ in (\ref{wg45}) as follows:
\begin{equation}\label{wg44}
	\begin{aligned}
		&\left( \tau ^t \right) ^2\mathbb{E} \left[ \left\| \boldsymbol{\bar{g}}^t \right\| _{2}^{2} \right] 
		\\
		=&\left( \tau ^t \right) ^2\mathbb{E} \left[ \left\| \sum_{n\in \mathcal{K} ^t}{\frac{p_{n,B}^{t}}{\sum_{n\in \mathcal{K} ^t}{p_{n,B}^{t}}}\nabla L_n\left( \boldsymbol{m}_{n}^{t} \right)} \right\| _{2}^{2} \right] 
		\\
		\overset{\left( a \right)}{\leqslant}&\left( \tau ^t \right) ^2\sum_{n\in \mathcal{K} ^t}{\frac{p_{n,B}^{t}}{\sum_{n\in \mathcal{K} ^t}{p_{n,B}^{t}}}\mathbb{E} \left[ \left\| \nabla L_n\left( \boldsymbol{m}_{n}^{t} \right) \right\| _{2}^{2} \right]}
		\\
		\leqslant& 2\theta \left( \tau ^t \right) ^2\sum_{n\in \mathcal{K} ^t}{\frac{p_{n,B}^{t}}{\sum_{n\in \mathcal{K} ^t}{p_{n,B}^{t}}}\mathbb{E} \left[ L_n\left( \boldsymbol{m}_{n}^{t} \right) -L_n\left( \boldsymbol{m}_{n}^{*} \right) \right]},
	\end{aligned}
\end{equation}
where (a) is obtained by applying Jensen's inequality and we applied the property of $\theta$-smooth function that
\begin{equation}\label{wg520}
	\begin{aligned}
		\left\| \nabla L_n\left( \boldsymbol{m}_{n}^{t} \right) \right\| _{2}^{2}\leqslant 2\theta \left[ L\left( \boldsymbol{m}_{n}^{t} \right) -L\left( \boldsymbol{m}_{n}^{*} \right) \right] 
	\end{aligned}
\end{equation}
in the last inequality. Then, the upper bound of term $B$ in (\ref{wg45}) can be given by
\begin{equation}\label{wg51}
	\begin{aligned}
		&-2\tau ^t\left< \boldsymbol{m}^t-\boldsymbol{m}^*,\boldsymbol{\bar{g}}^t \right> 
		\\
		=&2\tau ^t\sum_{n\in \mathcal{K} ^t}{\frac{p_{n,B}^{t}}{\sum_{n\in \mathcal{K} ^t}{p_{n,B}^{t}}}\mathbb{E} \left[ \left< \boldsymbol{m}^*-\boldsymbol{m}_{n}^{t},\nabla L_n\left( \boldsymbol{m}^t \right) \right> \right]}
		\\
		\overset{\left( a \right)}{\leqslant}&2\tau ^t\sum_{n\in \mathcal{K} ^t}{\frac{p_{n,B}^{t}}{\sum_{n\in \mathcal{K} ^t}{p_{n,B}^{t}}}\mathbb{E} \left[ L_n\left( \boldsymbol{m}^* \right) -L_n\left( \boldsymbol{m}_{n}^{t} \right) \right.}\left. -\frac{\rho}{2}\left\| \boldsymbol{m}^t-\boldsymbol{m}^* \right\| _{2}^{2} \right] ,
	\end{aligned}
\end{equation}
where (a) is from Assumption \ref{assumption2}. By combining (\ref{wg44}) with (\ref{wg51}), we obtain the upper bound of term $A+B$ as follows:
\begin{equation}\label{wg53}
	\begin{aligned}
		&\left( \tau ^t \right) ^2\mathbb{E} \left[ \left\| \boldsymbol{\bar{g}}^t \right\| _{2}^{2} \right] -2\tau ^t\left< \boldsymbol{m}^t-\boldsymbol{m}^*,\boldsymbol{\bar{g}}^t \right> 
		\\
		=&2\theta \left( \tau ^t \right) ^2\sum_{n\in \mathcal{K} ^t}{\frac{p_{n,B}^{t}}{\sum_{n\in \mathcal{K} ^t}{p_{n,B}^{t}}}\mathbb{E} \left[ L_n\left( \boldsymbol{m}_{n}^{t} \right) -L_n\left( \boldsymbol{m}_{n}^{*} \right) \right]}
		\\
		&-2\tau ^t\sum_{n\in \mathcal{K} ^t}{\frac{p_{n,B}^{t}}{\sum_{n\in \mathcal{K} ^t}{p_{n,B}^{t}}}\mathbb{E} \left[ L_n\left( \boldsymbol{m}_{n}^{t} \right) -L_n\left( \boldsymbol{m}^* \right) \right]}
		\\
		&-\rho \tau ^t\mathbb{E} \left[ \left\| \boldsymbol{m}^t-\boldsymbol{m}^* \right\| _{2}^{2} \right] 
		\\
		=&-2\tau ^t\left( 1-\theta \tau ^t \right) \sum_{n\in \mathcal{K} ^t}{\frac{p_{n,B}^{t}}{\sum_{n\in \mathcal{K} ^t}{p_{n,B}^{t}}}\mathbb{E} \left[ L_n\left( \boldsymbol{m}_{n}^{t} \right) -L_n\left( \boldsymbol{m}^* \right) \right]}
		\\
		&+2\theta \left( \tau ^t \right) ^2\sum_{n\in \mathcal{K} ^t}{\frac{p_{n,B}^{t}}{\sum_{n\in \mathcal{K} ^t}{p_{n,B}^{t}}}\mathbb{E} \left[ L_n\left( \boldsymbol{m}^* \right) -L_n\left( \boldsymbol{m}_{n}^{*} \right) \right]}-\rho \tau ^t\mathbb{E} \left[ \left\| \boldsymbol{m}^t-\boldsymbol{m}^* \right\| _{2}^{2} \right] 
		\\
		&\overset{\left( a \right)}{=}\underset{D}{\underbrace{-2\tau ^t\left( 1-\theta \tau ^t \right) \sum_{n\in \mathcal{K} ^t}{\frac{p_{n,B}^{t}}{\sum_{n\in \mathcal{K} ^t}{p_{n,B}^{t}}}\mathbb{E} \left[ L_n\left( \boldsymbol{m}_{n}^{t} \right) -L_n\left( \boldsymbol{m}^* \right) \right]}}}
		\\
		&+2\theta \left( \tau ^t \right) ^2\varGamma -\rho \tau ^t\mathbb{E} \left[ \left\| \boldsymbol{m}^t-\boldsymbol{m}^* \right\| _{2}^{2} \right] ,
	\end{aligned}
\end{equation}
where (a) is from Lemma \ref{lemma5}. To obtain the upper bound of term $D$ in (\ref{wg53}), we have
\begin{equation}\label{wg533}
	\begin{aligned}
		&-2\tau ^t\left( 1-\theta \tau ^t \right) \sum_{n\in \mathcal{K} ^t}{\frac{p_{n,B}^{t}}{\sum_{n\in \mathcal{K} ^t}{p_{n,B}^{t}}}\mathbb{E} \left[ L_n\left( \boldsymbol{m}_{n}^{t} \right) -L_n\left( \boldsymbol{m}^* \right) \right]}
		\\
		=&2\tau ^t\left( 1-\theta \tau ^t \right) \sum_{n\in \mathcal{K} ^t}{\frac{p_{n,B}^{t}}{\sum_{n\in \mathcal{K} ^t}{p_{n,B}^{t}}}\left[ \mathbb{E} \left[ L_n\left( \boldsymbol{m}_{n}^{*} \right) -L_n\left( \boldsymbol{m}_{n}^{t} \right) \right] \right.}
		\\
		&\left. +\mathbb{E} \left[ L_n\left( \boldsymbol{m}^* \right) -L_n\left( \boldsymbol{m}_{n}^{*} \right) \right] \right] 
		\\
		\overset{\left( a \right)}{\leqslant}&2\tau ^t\left( 1-\theta \tau ^t \right) \sum_{n\in \mathcal{K} ^t}{\frac{p_{n,B}^{t}}{\sum_{n\in \mathcal{K} ^t}{p_{n,B}^{t}}}\mathbb{E} \left[ L_n\left( \boldsymbol{m}^* \right) -L_n\left( \boldsymbol{m}_{n}^{*} \right) \right]}
		\\
		=&2\tau ^t\left( 1-\theta \tau ^t \right) \varGamma,
	\end{aligned}
\end{equation}
where (a) comes from that $2\tau ^t\left( 1-\theta \tau ^t \right) \geqslant 0$ and $L_n\left( \boldsymbol{m}_{n}^{*} \right) -L_n\left( \boldsymbol{m}_{n}^{t} \right) \leqslant 0$. Substituting (\ref{wg533}) back into (\ref{wg53}), we finally get the upper bound of term $A+B$ in (\ref{wg45}) as follows:
\begin{equation}\label{wwg51}
	\begin{aligned}
		&\left( \tau ^t \right) ^2\mathbb{E} \left[ \left\| \boldsymbol{\bar{g}}^t \right\| _{2}^{2} \right] -2\tau ^t\left< \boldsymbol{m}^t-\boldsymbol{m}^*,\boldsymbol{\bar{g}}^t \right> 
		\\
		\leqslant& 2\tau ^t\left( 1-\theta \tau ^t \right) \varGamma +2\theta \left( \tau ^t \right) ^2\varGamma -\rho \tau ^t\mathbb{E} \left[ \left\| \boldsymbol{m}^t-\boldsymbol{m}^* \right\| _{2}^{2} \right] 
		\\
		\leqslant& 2\tau ^t\varGamma -\rho \tau ^t\mathbb{E} \left[ \left\| \boldsymbol{m}^t-\boldsymbol{m}^* \right\| _{2}^{2} \right] 
		\\
		\overset{\left( a \right)}{\leqslant}&2\varrho \left( \tau ^t \right) ^2\varGamma-\rho \tau ^t\mathbb{E} \left[ \left\| \boldsymbol{m}^t-\boldsymbol{m}^* \right\| _{2}^{2} \right],
	\end{aligned}
\end{equation}
where (a) is from $\frac{1}{\varrho}\,\,\leqslant \tau ^t$. Then, the upper bound of term $C$ in (\ref{wg45}) is given by
\begin{equation}\label{wwg52}
	\begin{aligned}
		&\mathbb{E} \left[ \left\| \boldsymbol{r}_{B,Tot}^{t} \right\| _{2}^{2} \right] 
		\\
		=&\mathbb{E} \left[ \left\| \sum_{n\in \mathcal{J} ^t}{\frac{p_{n,B}^{t}}{\sqrt{d}}\boldsymbol{e}_{n}^{t}}+\boldsymbol{r}_{B}^{t} \right\| _{2}^{2} \right] 
		\\
		\overset{\left( a \right)}{\leqslant}&\mathbb{E} \left[ \left\| \sum_{n\in \mathcal{J} ^t}{\frac{p_{n,B}^{t}}{\sqrt{d}}\boldsymbol{e}_{n}^{t}} \right\| _{2}^{2} \right] +\mathbb{E} \left[ \left\| \boldsymbol{r}_{B}^{t} \right\| _{2}^{2} \right] 
		\\
		\overset{\left( b \right)}{\leqslant}& \left| \mathcal{J} ^t \right|\sum_{n\in \mathcal{J} ^t}{\frac{\left( p_{n,B}^{t} \right) ^2}{d}\mathbb{E} \left[ \left\| \boldsymbol{e}_{n}^{t} \right\| _{2}^{2} \right]}+d\sigma _B
		\\
		\leqslant& N\sum_{n\in \mathcal{J} ^t}{\left( p_{n,B}^{t} \right) ^2}+d\sigma _B,
	\end{aligned}
\end{equation}
where (a) is from the fact that $\mathbb{E} \left[ \boldsymbol{e}_{n}^{t} \right] =\mathbb{E} \left[ \boldsymbol{r}_{B}^{t} \right] =0
$ and step (b) is obtained by applying Jensen's inequality.

Finally, by substituting (\ref{wwg51}), (\ref{wwg52}) into (\ref{wg45}), we get the upper bound of $\mathbb{E} \left[ \left\| \boldsymbol{m}^{t+1}-\boldsymbol{m}^* \right\| _{2}^{2} \right]$ as follows:
\begin{equation}\label{wg57}
	\begin{aligned}
		\mathbb{E} \left[ \left\| \boldsymbol{m}^{t+1}-\boldsymbol{m}^* \right\| _{2}^{2} \right] &\leqslant \left( 1-\rho \tau ^t \right) \mathbb{E} \left[ \left\| \boldsymbol{m}^t-\boldsymbol{m}^* \right\| _{2}^{2} \right] 
		\\
		&+2\varrho \left( \tau ^t \right) ^2\varGamma +\left( \tau ^t \right) ^2\vartheta ^2+\left( \frac{\tau ^tG}{\sum_{n\in \mathcal{K} ^t}{p_{n,B}^{t}}} \right) ^2\left( N\sum_{n\in \mathcal{J} ^t}{\left( p_{n,B}^{t} \right) ^2}+d\sigma _B \right) .
	\end{aligned}
\end{equation}
Then, we complete the proof of Theorem 1.

\section{Proof of Corollary 1}\label{proofofcorollary1}
Similar to \cite{yan2022performance}, we define, 
\begin{equation}\label{wg57}
	\begin{aligned}
		\varOmega ^t=\mathbb{E} \left[ \left\| \boldsymbol{m}^t-\boldsymbol{m}^* \right\| _{2}^{2} \right].	
	\end{aligned}
\end{equation}

It thus follows from Theorem \ref{lemmaconvergenceanalysis} that 
\begin{equation}\label{wg57}
	\begin{aligned}
		\varOmega ^{t+1}\leqslant \left( 1-\rho \tau ^t \right) \varOmega ^t+\left( \tau ^t \right) ^2\eta ^t,
	\end{aligned}
\end{equation}
where $\eta ^t=2\varrho \varGamma +\vartheta ^2+G^2\varPsi ^t.$

Let $\tau ^t=\frac{\lambda}{t+\mu}$ for some $\lambda \geqslant \frac{1}{\rho}$ and $\mu >1$ such that $\tau ^0\leqslant \left\{ \frac{1}{\rho},\frac{1}{\theta} \right\} =\frac{1}{\theta}$. We will prove
\begin{equation}\label{wg57}
	\begin{aligned}
		\varOmega ^t\leqslant \frac{\chi}{t+\mu},
	\end{aligned}
\end{equation} 
where $\chi =\max \left\{ \frac{\lambda ^2\eta}{\lambda \rho -1},\mu \varOmega ^0 \right\}$ with $\eta =\underset{t}{\max}\left\{ \eta ^t \right\}$ by induction as follows:

Firstly, the inequality naturally holds for $t = 0$ according to the definition of $\chi$.

Then, assuming that the inequality holds for $t>0$, it follows that,

\begin{equation}\label{wg57}
	\begin{aligned}
		&\varOmega ^{t+1}\leqslant \left( 1-\rho \tau ^t \right) \varOmega ^t+\left( \tau ^t \right) ^2\eta ^t
		\\
		=&\left( 1-\frac{\lambda \rho}{t+\mu} \right) \frac{\chi}{t+\mu}+\frac{\lambda ^2\eta ^t}{\left( t+\mu \right) ^2}
		\\
		=&\frac{t+\mu -1}{\left( t+\mu \right) ^2}\chi +\underset{\leqslant 0}{\underbrace{\frac{\lambda ^2\eta ^t}{\left( t+\mu \right) ^2}-\frac{\lambda \rho -1}{\left( t+\mu \right) ^2}\chi }}
		\\
		\leqslant& \frac{t+\mu -1}{\left( t+\mu \right) ^2-1}\chi =\frac{\chi}{t+1+\mu}.
	\end{aligned}
\end{equation}
Specifically, we choose $\lambda =\frac{2}{\rho}$, $\mu =\frac{2\theta}{\rho}$, and then $\tau ^t=\frac{2}{\rho t+2\theta}$. Let $\frac{1}{\varrho}=\tau ^T=\frac{2}{\rho T+2\theta}$ so that $\frac{1}{\varrho}\leqslant \tau ^t\leqslant \frac{1}{\theta}$, then, one has
\begin{equation}\label{wg57}
	\begin{aligned}
		&\mathbb{E} \left[ L\left( \boldsymbol{m}^T \right) \right] -L^*
		\\
		\overset{\left( a \right)}{\leqslant}&\frac{\theta}{2}\mathbb{E} \left[ \left\| \boldsymbol{m}^T-\boldsymbol{m}^* \right\| _{2}^{2} \right] 
		\\
		\overset{\left( b \right)}{\leqslant}&\frac{\theta}{2}\left( \frac{\chi}{T+\mu} \right) 
		\\
		\leqslant& \frac{\theta}{2\left( T+\mu \right)}\max \left\{ \frac{\lambda ^2\eta}{\lambda \rho -1},\mu \varOmega ^0 \right\} 
		\\
		\leqslant& \frac{\rho \theta}{2\left( \rho T+2\theta \right)}\left( \frac{4\eta}{\rho ^2}+\frac{2\theta}{\rho}\left\| \boldsymbol{m}^0-\boldsymbol{m}^* \right\| _{2}^{2} \right) 
		\\
		=&\frac{\theta}{\rho T+2\theta}\left( \frac{2\eta}{\rho}+\theta \left\| \boldsymbol{m}^0-\boldsymbol{m}^* \right\| _{2}^{2} \right) 
		\\
		=&\frac{\theta}{\rho T+2\theta}\left[ \frac{2}{\rho}\left( \vartheta ^2+G^2\underset{t}{\max}\left\{ \varPsi ^t \right\} \right) \right] +\frac{2\varGamma \theta}{\rho},
	\end{aligned}
\end{equation}
where (a) is from Lemma \ref{lemma1} and the fact that $\nabla L\left( \boldsymbol{m}^* \right) =0$. Then, we complete the proof of Corollary 1.

\bibliographystyle{IEEEtran}
\bibliography{FLbib}

\end{document}